\documentclass[
preprint,
numerical, 
showkeys
]{jasatex}

\usepackage{amsmath}
\usepackage{amsfonts}
\usepackage{amssymb}
\usepackage{graphicx}
\usepackage{float}
\usepackage[titletoc]{appendix}
\usepackage{fullpage}
\usepackage[stable]{footmisc}

\newcommand{\br}[1]{\left( #1 \right)}
\newcommand{\sbr}[1]{\left[ #1 \right]}

\newcommand{\ex}{\mathrm{e}}
\newcommand{\ii}{\mathrm{i}}

\newcommand{\abs}[1]{\left| #1 \right|}

\newcommand{\bk}{\boldsymbol{k}}

\newcommand{\pa}[1]{\partial_{#1}}
\newcommand{\Ja}[2]{\mathcal{J}\left(#1,#2\right)}

\newcommand{\dd}{\mathrm{d}}

\newcommand{\Rr}{\mathrm{R}_\rho}
\newcommand{\Ra}{\mathrm{Ra}}
\newcommand{\Sc}{\mathrm{Sc}}
\newcommand{\Nu}{\mathrm{Nu}}
\newcommand{\R}{\mathcal{R}}
\newcommand{\half}{1/2}

\begin{document}

\preprint{}

\title[A reduced model for salt-finger convection]{A reduced model for salt-finger convection\\ in the small diffusivity ratio limit}

\author{Jin-Han Xie$^{1}$}%
\email{j.h.xie@berkeley.edu}
\author{Benjamin Miquel$^{2}$}%
\email{benjamin.miquel@colorado.edu}
\author{Keith Julien$^{2}$}%
\email{julien@colorado.edu}
\author{Edgar Knobloch$^{1,}$*}%
\email{knobloch@physics.berkeley.edu}

\affiliation{%
	$^{1}$ Department of Physics, University of California, Berkeley, CA 94720, USA\\
	$^{2}$ Department of Applied Mathematics, University of Colorado at Boulder, Boulder, CO 80309, USA}

\begin{abstract}
A simple model of nonlinear salt-finger convection in two dimensions is derived and studied. The model is valid in the limit of small solute to heat diffusivity ratio and large density ratio, which is relevant to both oceanographic and astrophysical applications. Two limits distinguished by the magnitude of the Schmidt number are found. For order one Schmidt numbers, appropriate for astrophysical applications, a modified Rayleigh-B\'enard system with large-scale damping due to a stabilizing temperature is obtained. For large Schmidt numbers, appropriate for the oceanic setting, the model combines a prognostic equation for the solute field and a diagnostic equation for inertia-free momentum dynamics. Two distinct saturation regimes are identified for the second model: The weakly driven regime is characterized by a large-scale flow associated with a balance between advection and linear instability, while the strongly driven regime produces multiscale structures, resulting in a balance between the energy input through linear instability and the energy transfer between scales. For both regimes, we analytically predict and numerically confirm the dependence of the kinetic energy and salinity fluxes on the ratio between solute and heat Rayleigh numbers. The spectra and probability density functions are also computed.
\end{abstract}


\maketitle



\section{Introduction}

Doubly diffusive systems in which two components with different diffusivities contribute to buoyancy in opposite ways arise frequently in geophysics and astrophysics \cite{Turn1974,Schm1994}: for example, heat and salt contribute to the stratification in the oceanic setting, heat and chemical composition in the stellar astrophysics, and sugar and salt in laboratory experiments. In the following, motivated by oceanic dynamics, we refer to the slowly diffusing component as salinity and to the rapidly diffusing component as temperature. Under appropriate conditions, the different diffusivities of these two components may destabilize an otherwise stably stratified configuration and hence lead to mixing. Two scenarios are possible. In the first, temperature stratification is stabilizing but the salinity stratification is destabilizing, a configuration that leads to a fingering instability. In the second scenario, where the salinity stratification is stabilizing while the temperature is destabilizing, a diffusive oscillatory instability called overstability may take place. In both these cases instability is present because heat diffuses more rapidly than a typical solute, i.e. the ratio $\tau$ of the solute diffusivity to thermal diffusivity (or inverse Lewis number) satisfies $\tau<1$. In this paper we concentrate on the nonlinear properties of the turbulent state arising in the former case, and suppose that the destabilizing effect is provided by an unstable salt stratification as frequently occurs in the oceans, focusing on the case $\tau\ll 1$.

Linear stability properties of the salt-finger regime were first determined by \citet{Ster1960}. As the magnitude of the unstable mode grows secondary instabilities are triggered \citep{Holy1984}, and finally lead to a statistically steady state \citep{Shen1995,Radk2010}. However, the physical processes behind the presence of such an equilibrium state are still incompletely understood. Stern \cite{Ster1969} has suggested that a collective instability can lead to the disruption of the salt-finger field and lead to saturation, and identified a dimensionless quantity, now called the Stern number, that can be used as a criterion for saturation.  Others have supposed that saturation will arise once the growth rate of a secondary instability, such as that identified in \cite{Holy1984}, becomes comparable to that of the salt-finger instability itself \cite{Radk2012,Brown2013}. More recently \citet{Shen1995} has suggested that finger collision plays an essential role because of the generation of large vertical gradients and hence strong dissipation. In an interesting paper \citet{Paparella2012} proposed a saturation mechanism based on finger clustering that leads to the generation of large-scale structures and hence to saturation at the large scale.

Except for weakly nonlinear approaches, eg. \cite{Proc1986,Radk2010}, most researchers employ numerical studies of the primitive equations consisting of the Navier-Stokes equation coupled to scalar equations for heat and salinity transport. Such studies, particularly in three dimensions, have proved invaluable and identified a number of novel processes, eg. \cite{Trax2011}. To shed light on some of these we propose here a systematic procedure that leads to a simplified set of equations that are easier to study, both theoretically and numerically. In the following we refer to these, following  \cite{Juli2007}, as {\it reduced equations}. The procedure described below has been used before \cite{SternRadko1998,RadkoStern1999,Prat2015} and focuses on the double limit of a large density ratio, where the stratification is dominated by thermal contributions, and an asymptotically small diffusivity ratio, where salinity diffuses much more slowly than heat. The latter is motivated by the small value of the diffusivity ratio in the ocean while large density ratios are observed in fingering layers in experiments \cite{Schmitt1979} and in some oceanic measurements \cite{StLaurent1999}, and is a limit beyond the current capability of direct numerical simulations. Two reduced models are derived depending on the order of Schmidt number $\Sc$, which captures the relative strength of momentum and salinity diffusion. When these two effects are comparable, $\Sc=O(1)$, a reduced model that bears similarity to the Rayleigh-Bénard configuration is derived. This model is applicable to astrophysical objects, eg. \cite{Traxler2011,Brown2013,Prat2015}. In the oceanic parameter regime with fast momentum diffusion, $\Sc \gg 1$, we derive a reduced model with a prognostic-diagnostic form, which we refer to as the inertia-free salt convection model. This model is the main topic of this paper.

It turns out that this model already appears in the work of Radko and Stern \cite{RadkoStern1999}. However, beyond one three-dimensional simulation in a very small aspect ratio domain, these authors did not investigate the properties of this model. We believe in fact that the model merits far greater attention and that it correctly captures the saturation of the salt-finger instability and the resulting transport properties in a physically relevant regime. In this paper we therefore study the properties of this model at some length albeit in two dimensions. To do so we employ a doubly periodic setting as appropriate for oceanic and astrophysical applications with distant boundaries, in contrast to earlier studies of salt-finger convection in vertically bounded domains, eg. \cite{Howard1987,Radko2000,Yang2015,Yang2015JFM,Yang2016}. In this formulation elevator modes consisting of vertically invariant spatial dynamics are exact nonlinear solutions and we believe that analogous modes are crucial even in vertically bounded domains because of their efficiency in extracting energy from the salinity field.


The structure of this paper is as follows. In \S \ref{Sec_Formulation} we formulate the salt-finger convection problem followed in \S \ref{Sec_Smalldiff} by a derivation of two distinct reduced models valid for finite and infinite Schmidt numbers, respectively. The time evolution and the properties of the saturated state of the latter are studied in \S \ref{Sec_Evol} and \S \ref{Sec_SatSta} through a combination of numerical simulation and scaling analysis. Our results are summarized in \S \ref{Sec_Dis}. Two appendices, one summarizing the dependence of the results on the domain size and the other an analytical approximation to one of the saturation regimes, complete the paper.

\section{Formulation} \label{Sec_Formulation}

\subsection{Setup and nondimensionalization}

We consider a two-dimensional Boussinesq fluid of infinite extent in the horizontal and vertical directions, denoted by $x$ and $z$, respectively. An initial uniform density stratification is generated by means of a stabilizing background temperature gradient $\beta_T>0$ and a destabilizing background salinity gradient $\beta_S>0$. We introduce the thermal and saline expansion coefficients $\alpha_T>0$ and $\alpha_S>0$ so that the density of the fluid $\rho(T,S)$ at temperature $T$ and salinity $S$ is given by $\rho(T,S) = \rho_0(1-\alpha_T(T-T_0) + \alpha_S (S-S_0))$, where $\rho_0=\rho (T_0,S_0)$ is a reference state. The remaining material properties of the fluid are specified by the thermal diffusivity $\kappa_T$, the solute diffusivity $\kappa_S$ and the viscosity $\nu$, all of which are taken to be constant. The acceleration due to gravity is denoted by $g$.

The equations are made dimensionless using the natural salt finger width $d$ given by
\begin{equation}
d = \left(\frac{\nu\kappa_T}{g\alpha_T \beta_T}\right)^{1/4}\, .
\end{equation}
This length defines the characteristic temperature $\mathcal{T}$ and salinity $\mathcal{S}$ scales:
\begin{subequations}
\begin{align}
\mathcal{T} &= \beta_T d\, , \\
\mathcal{S} &= \beta_S d\, .
\end{align}
\end{subequations}
These scales are used to nondimensionalize the temperature and salinity profiles $T_{\mathrm{total}}$ and $S_{\mathrm{total}}$ which are decomposed into the initial background state and a fluctuating component indicated by a tilde:
\begin{subequations}
\begin{gather}
T_{\mathrm{total}} = z + \tilde{T}\, ,\\
S_{\mathrm{total}} = z + \tilde{S}\, .
\end{gather}
\end{subequations}
The incompressible velocity field is expressed in terms of a streamfunction $\psi$,
\begin{equation}
\boldsymbol{u} \equiv u \boldsymbol{e}_x + w\, \boldsymbol{e}_z = \begin{pmatrix}
-\partial_z \psi \\ \partial_x \psi\, ,
\end{pmatrix}  
\end{equation} where $(u,w)$ and $\psi$ are made dimensionless using $\kappa_S/d$ and $\kappa_S$, respectively. Finally, time is made dimensionless using the salinity diffusion timescale:
\begin{equation}
t_{\mathrm{salt}} = \frac{d^2}{\kappa_S}\, .
\end{equation}

\subsection{Nondimensional equations}
The governing dimensionless equations for fluctuations around the conduction state $\psi=0$, $T_{\mathrm{total}} = S_{\mathrm{total}} = z$ are
\begin{subequations}\label{GovEq}
	\begin{align}
	\frac{\tau}{\Pr}\sbr{ \frac{\partial }{\partial t} \nabla^2 \psi + \Ja{\psi}{\nabla^2\psi} } &= \frac{1}{\tau}\frac{\partial \tilde T}{\partial x} - \frac{1}{\tau \Rr}\frac{\partial \tilde S}{\partial x} + \nabla^4 \psi,\\
	\frac{\partial }{\partial t} \tilde T + \Ja{\psi}{\tilde T} + \frac{\partial \psi}{\partial x} &= \tau^{-1} \nabla^2 \tilde T,\label{TEq}\\
	\frac{\partial }{\partial t}\tilde S + \Ja{\psi}{\tilde S} + \frac{\partial \psi}{\partial x} &=  \nabla^2 \tilde S. \label{SEq}
	\end{align}
\end{subequations}
Three dimensionless ratios enter this set of equations: the Prandtl number $\Pr$, the inverse Lewis number $\tau$ and the density ratio $\Rr$ specifying of the relative contribution of the temperature and salinity to the background density gradient. A fourth dimensionless ratio, the Schmidt number $\mathrm{Sc}$, can be introduced as an alternative to the diffusivity ratio. These quantities are defined by
\begin{equation}
\Pr = \frac{\nu}{\kappa_T},\qquad \tau =\frac{ \kappa_S }{\kappa_T},\qquad \mathrm{Sc} = \frac{\nu}{\kappa_S},\qquad \Rr = \frac{\alpha_T\beta_T}{\alpha_S\beta_S}\,.
\end{equation}
In this paper, we adopt doubly periodic boundary conditions for $\psi$, $\tilde T$ and $\tilde S$, and consider a statically stable configuration with $\Rr >1$.

\subsection{Connection with an alternative description in terms of flux Rayleigh numbers}
The above formulation can be connected to the classical description of doubly diffusive convection in terms of the temperature and salinity Rayleigh numbers, $\Ra_T$ and $\Ra_S$, respectively. These are defined for a layer of vertical extent $H$, viz.,
\begin{equation}
\Ra_T = \frac{g\alpha_T\beta_T H^4}{\nu \kappa_T} \quad \mathrm{and} \quad \Ra_S = \frac{g\alpha_S\beta_S H^4}{\nu \kappa_S}\, ,
\end{equation}
implying that
\begin{equation}
\Rr \tau = \frac{\Ra_T}{\Ra_S}\, .
\end{equation}
Large $\Ra_T$ flows correspond to slender fingers (at least in the linear regime) as the aspect ratio $d/H$ of the fingers is given by 
\begin{equation}
  \frac{d}{H}=\Ra_T^{-1/4}. \label{RaT}  
\end{equation}

\subsection{Linear stability analysis}
The growth of salt fingers can be understood by analyzing the linear stability of the conduction state with respect to normal mode perturbations of the form $\br{\psi(x,z,t), \tilde T(x,z,t), \tilde S(x,z,t)}= \Re \left\{ (\psi_e, T_e, S_e) \ex^{\lambda t + \ii (kx+mz)} \right\}$. The growth rate $\lambda$ obeys the dispersion relation
\begin{multline}
\lambda^3 +|K|^2\br{1+\frac{1}{\tau}+\frac{\Pr}{\tau}} \lambda^2 + \\ \sbr{ |K|^4\br{\frac{1}{\tau}+\frac{\Pr}{\tau}+
\frac{\Pr}{\tau^2}} + \frac{\Pr k^2}{\tau^2 |K|^2}\br{1-\frac{1}{\Rr}} } \lambda + \frac{\Pr}{\tau^2} \sbr{ |K|^6+ k^2\br{1-\frac{1}{\tau \Rr} }} = 0\, ,
\end{multline}
where $|K|^2\equiv k^2+m^2$. For a Rayleigh-Taylor stable layer such that $\Rr>1$, the cubic, quadratic and linear coefficients of this dispersion relation are positive. An exponentially growing instability is present if and only if there exists a wavenumber $k$ such that the constant term is negative. This condition is equivalent to
\begin{equation}\label{critical_Ra_fullset}
\frac{1}{\tau \Rr} >1 \, .
\end{equation}
The salt-finger instability is therefore present in the regime
\begin{equation}
 1 < \Rr  <  \frac{1}{\tau}.
\end{equation} 
This condition defines the regime of interest for the remainder of this paper.

The strongest linear instability is associated with vertically invariant modes with $m=0$, called elevator modes in the literature. With periodic boundary conditions in the vertical, these modes are exact solutions of the system (\ref{GovEq}) and so play an important role in the dynamics. In a vertically bounded domain elevator modes are not permitted, but their effect persists, at least when the domain has a large vertical extent, $H\gg d$.

\section{Reduced models with small diffusivity ratio} \label{Sec_Smalldiff}
\subsection{Scalings of parameters and variables}\label{SecModDer}
In this section we focus on the small diffusivity ratio limit $\tau\ll 1$ of the system (\ref{GovEq}), i.e. $\kappa_S \ll \kappa_T$. Two cases are considered, distinguished by the magnitude of the Schmidt number $\mathrm{Sc}$. For Schmidt number of order one, where $\kappa_S\sim \nu $, we obtain the small Prandtl number regime $\mathrm{Pr}=O(\tau)$ of astrophysical relevance. This results in a modified Rayleigh-B\'enard system with salinity-driven instability and rapidly diffusing temperature, as described in \S \ref{Sec_MRBC}. The large Schmidt number regime, where $\kappa_S \ll \nu$ and $\mathrm{Pr} \gg \tau$ is relevant for oceanic thermohaline flows where $\mathrm{Sc}\approx 700$ and $\tau \approx 0.01$. 
This results in a model where inertial forces are small and salinity is the only slowly diffusing quantity
\begin{subequations}
	\begin{gather}
	\kappa_S \ll \kappa_T,\\ \kappa_S \ll \nu\, . \label{IFSC_regime}
	\end{gather}
\end{subequations}
We refer to this system, considered in \S \ref{Sec_IFSC}, as the inertia-free salt convection model.

In all cases we suppose, moreover, that the density ratio $\Rr$ is large and comparable to $\tau^{-1}$. 
This assumption is supported by both laboratory experiments in the fingering domain \cite{Linden1973} and oceanic measurements \cite{StLaurent1999} where $\Rr$ can reach $O(10^2)$. Also, in the astrophysics context, e.g. for stars along the red giant branch (RGB), $\Rr$ can be as large as $O(10^6)$, which is of the same order as the corresponding Prandtl number \cite{Denissenkov2010}.
Hence, we define the order one parameter:
\begin{equation}
\Ra = \frac{1}{\Rr \tau} = \frac{\Ra_S}{\Ra_T}\, . \label{DefRa}
\end{equation}
This parameter is analogous to a Rayleigh number for our system, as we demonstrate below. The parameter $\mathcal{R}$,
\begin{equation}
\mathcal{R} \equiv  \Ra -1 = \frac{1}{\Rr \tau} -1
\end{equation}
measures the supercriticality of the system and will therefore prove to be useful as well.

We now discuss the scaling of the fluid variables, starting with the streamfunction which remains untouched. The temperature fluctuations $\tilde{T}$ are scaled with $\tau$ and become an asymptotically small quantity in the limit $\tau \downarrow 0$.  In this limit the salinity fluctuations $\tilde{S}$ remain finite but are scaled for convenience by the order one factor $\mathrm{Ra}$:
\begin{subequations}
	\begin{gather}
	T_{\mathrm{total}} = z + \tau\, T\, ,\\
	S_{\mathrm{total}} = z + \Ra\, S\, ,
	\end{gather}
	\label{T_S_renormalized}
\end{subequations}
where $T$ and $S$ are now order one quantities. This formulation permits a feedback on the background salinity profile but excludes any leading order modification of the background temperature profile.

\subsection{Modified Rayleigh-B\'enard convection model: $\mathrm{Sc}= O(1)$}  \label{Sec_MRBC}
In the limit of small $\tau$ and $\mathrm{Sc} = O(1)$, with the scaling $\tilde T=\tau T$, we obtain a distinguished regime described by the following set of equations:
\begin{subequations}\label{mRBC}
	\begin{align}
	\frac{1}{\mathrm{Sc}}\sbr{ \frac{\partial }{\partial t} \nabla^2 \psi + \Ja{\psi}{\nabla^2\psi} } &= -\frac{\partial S}{\partial x} + \br{\nabla^4 + \Delta^{-1}\pa{x}^2} \psi\, ,\label{RBClike2} \\
	\frac{\partial }{\partial t} S + \Ja{\psi}{S} + \Ra\frac{\partial \psi}{\partial x} &= \nabla^2 S\, .\label{RBClike3} 
	\end{align}
\end{subequations}
Once $\psi$ is determined the scaled temperature $T$ is obtained from 
\begin{equation}
\frac{\partial \psi}{\partial x} = \nabla^2 T\, . \label{diag_temp}
\end{equation}
Equations~(\ref{mRBC}) describe a modified Rayleigh-B\'enard Convection (mRBC) system with large-scale vorticity dissipation given by $\Delta^{-1}\pa{x}^2 \psi$ in addition to the usual small-scale vorticity dissipation via $\nabla^4\psi$. This new source of dissipation is brought about by the stabilizing temperature field and serves to select an intrinsic finite-scale optimal mode, in contrast to the standard RBC system where the scale of the optimal mode is determined by the domain height. The resulting model applies under conditions prevailing in astrophysics, for instance for RGB stars, characterized by extremely small parameters $\Pr=O(\tau)\sim 10^{-6}$. At present, three-dimensional direct numerical simulations of the primitive doubly diffusive equations can only reach $\Pr=O(\tau)\sim 10^{-2}$ \cite{Traxler2011}.

\subsection{Inertia-free salt convection model: $Sc \gg 1$} \label{Sec_IFSC}

In this regime, characterized by condition (\ref{IFSC_regime}), viscosity is larger than the diffusivity of the salinity and as a result inertial modes are suppressed. This regime is relevant to the oceans and yields a further reduction of Eqs. (\ref{Sec_MRBC}) to a model which is now first order in time:
\begin{subequations}\label{RedMod}
	\begin{gather}
	\br{\pa{x}^2 + \nabla^6} \psi = \pa{x}\nabla^2 S\, , \label{Dia_Rel}\\
	\frac{\partial }{\partial t} S + \Ja{\psi}{S} + \Ra\frac{\partial \psi}{\partial x} = \nabla^2 S. \label{Pro_Rel}
	\end{gather}
\end{subequations}
In the following we refer to these equations as the inertia-free salt convection (IFSC) model. As before, the temperature $T$ is determined from equation (\ref{diag_temp}).

The IFSC model is the simplest model of salt-finger convection that can be derived and depends on the single intrinsic parameter $\Ra$ only. The coupled form of this model is reminiscent of other geophysical models such as those describing $\beta$-plane vorticity dynamics \cite{Rhines1975}, B\'enard convection in the large Prandtl number limit \cite{Busse1967} and $\beta$-convection \cite{Brummell1993}. However, the diagnostic equation~(\ref{Dia_Rel}) exhibits two distinguishing features: (i) in contrast to the aforementioned models, the salinity and streamfunction are necessarily out of phase, a crucial fact for linear instability; (ii) the existence of a complex salinity-streamfunction relation where the two terms $\pa{x}^2$ and $\nabla^6$ dominate on different horizontal scales. Consequently, (\ref{Dia_Rel}) implies the potential for dynamics on multiple spatial scales, a fact we believe provides the key to explaining the observed transitions in the saturated state as $\Ra$ increases, see \S \ref{Sec_Regimes}. 

In view of its simplicity and its strong relevance to oceanic flows, we focus in this paper on the IFSC model (\ref{RedMod}), and leave the mRBC model (\ref{mRBC}) for future study.

\section{IFSC model: from the linear instability to the saturated state} \label{Sec_Evol}

In this section we document a scenario that leads from vanishingly small perturbations of the purely conductive state to a convective saturated state in the IFSC model. During the first stage, a linear instability occurs and preferentially amplifies vertically invariant structures (elevator modes). These structures are then subject to a secondary instability resulting in undulations in the vertical direction. Finally the system saturates and exhibits structures of a much shorter vertical extent than observed during the first two phases. 

\subsection{Linear stability} \label{SecLinSta}
Substituting a normal mode ansatz of the form $\ex^{\lambda t+\ii(kx+mz)}$ into Eqs. (\ref{RedMod}) linearized around $S=\psi=0$ we obtain the growth rate $\lambda$ as a function of $\Ra$:
\begin{equation}
\lambda = \Ra \frac{k^2|K|^2}{k^2+|K|^6} - |K|^2, \label{GroRat}
\end{equation}  
where, as before, $|K|^2=k^2+m^2$. It follows that the threshold for instability, determined in Eq.~(\ref{critical_Ra_fullset}), is preserved in the IFSC model and occurs at
\begin{equation}
\Ra_c = 1\, .
\end{equation}
Thus the quantity $\mathcal{R} \equiv\Ra -1$ measures the distance from threshold, i.e., the supercriticality of the system.

\begin{figure}
	\centering
	\includegraphics[width=0.45\linewidth]{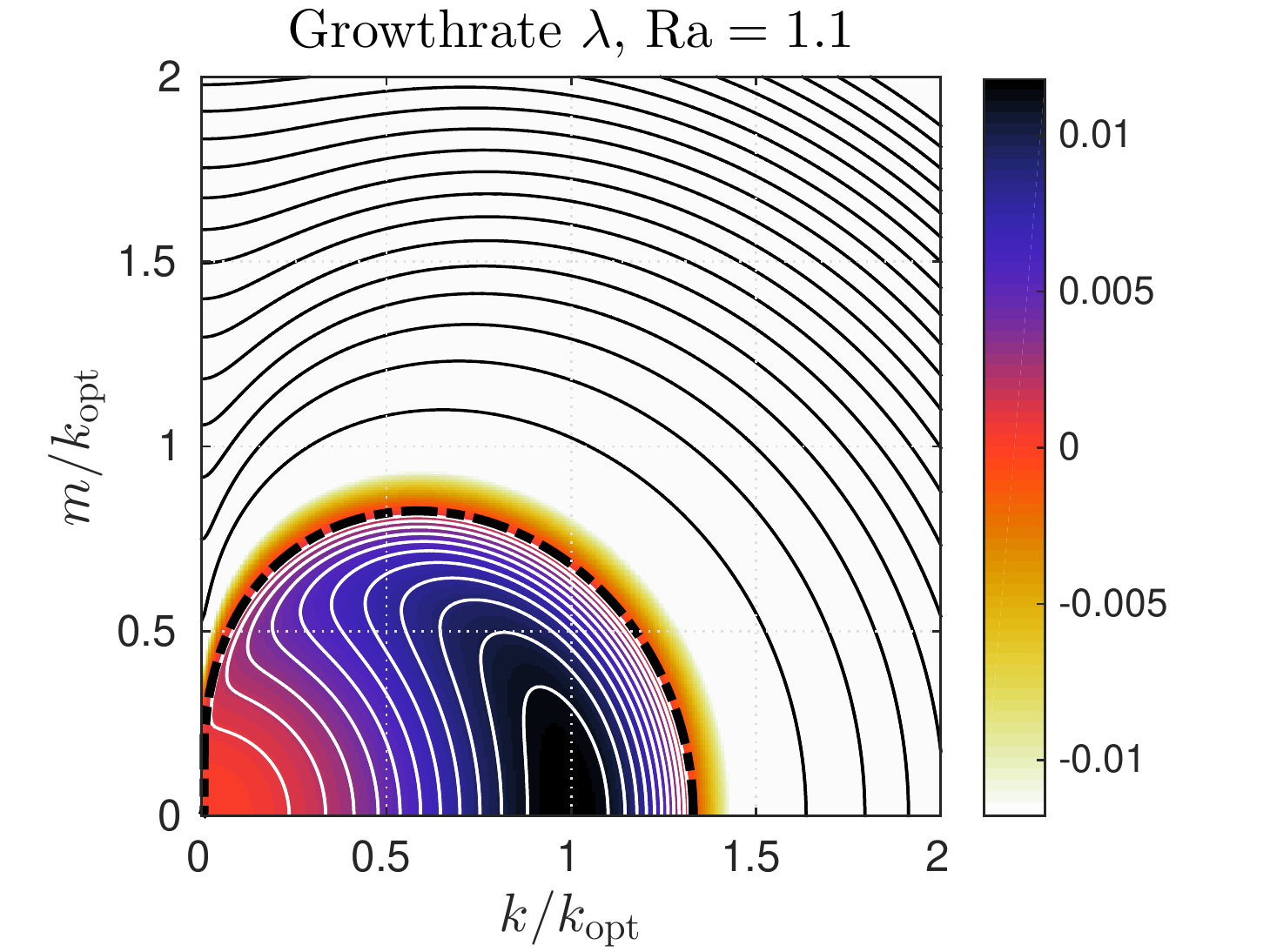}
	\includegraphics[width=0.45\linewidth]{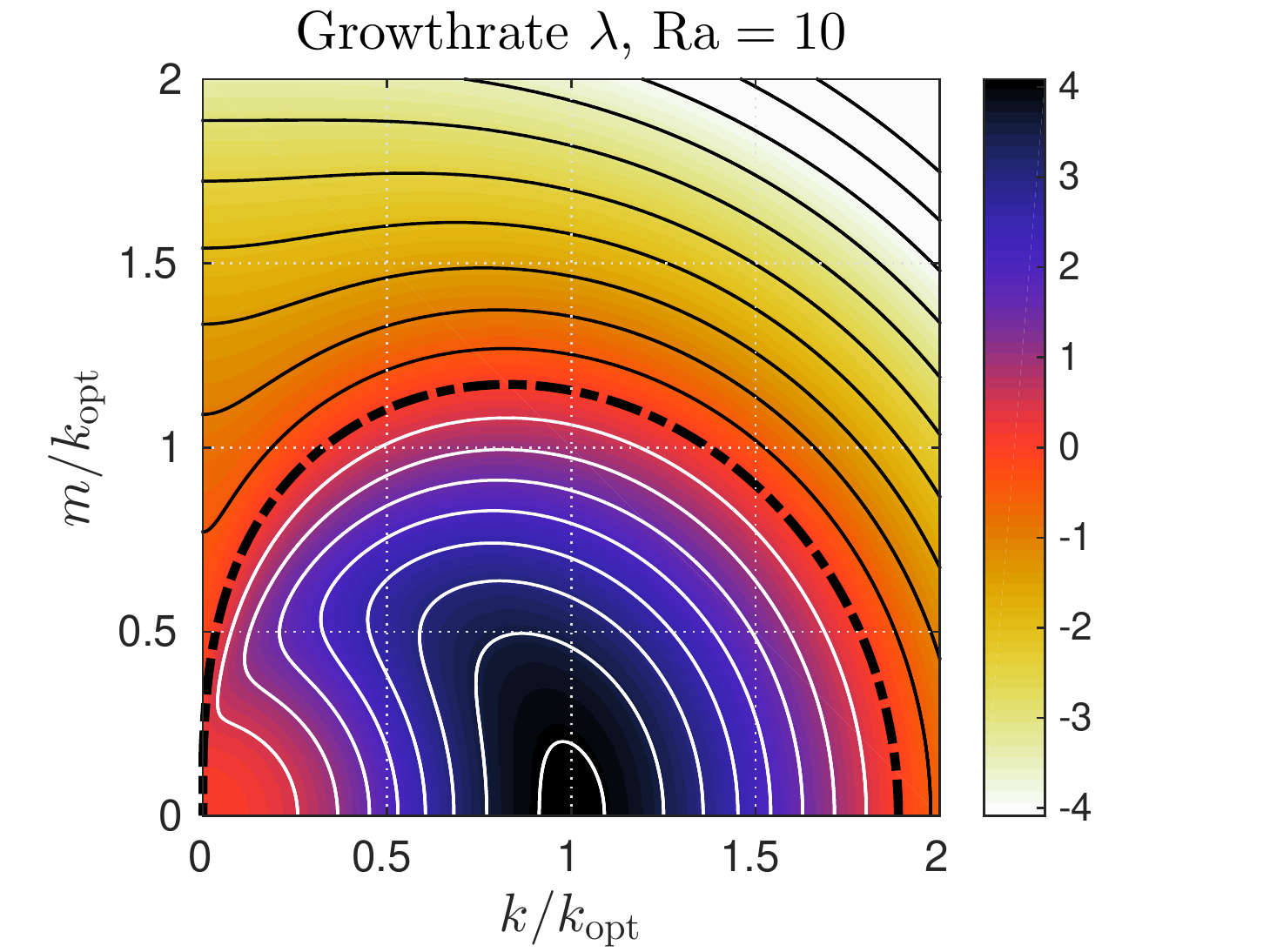}
	\caption{Contour plots of the linear growth rate $\lambda(k,m)$ of the reduced model (\ref{RedMod}). 
		The left panel shows the results for $\Ra=1.1$ with equispaced contours separated by 0.001 for positive values and 0.05 for negative values.
		The right panel shows the results for $\Ra=10$ with equispaced contours separated by 0.5.
	}
	\label{Linear_Growth_Rate_2D}
\end{figure}

\begin{figure}
	\centering
	\includegraphics[width=0.4\linewidth]{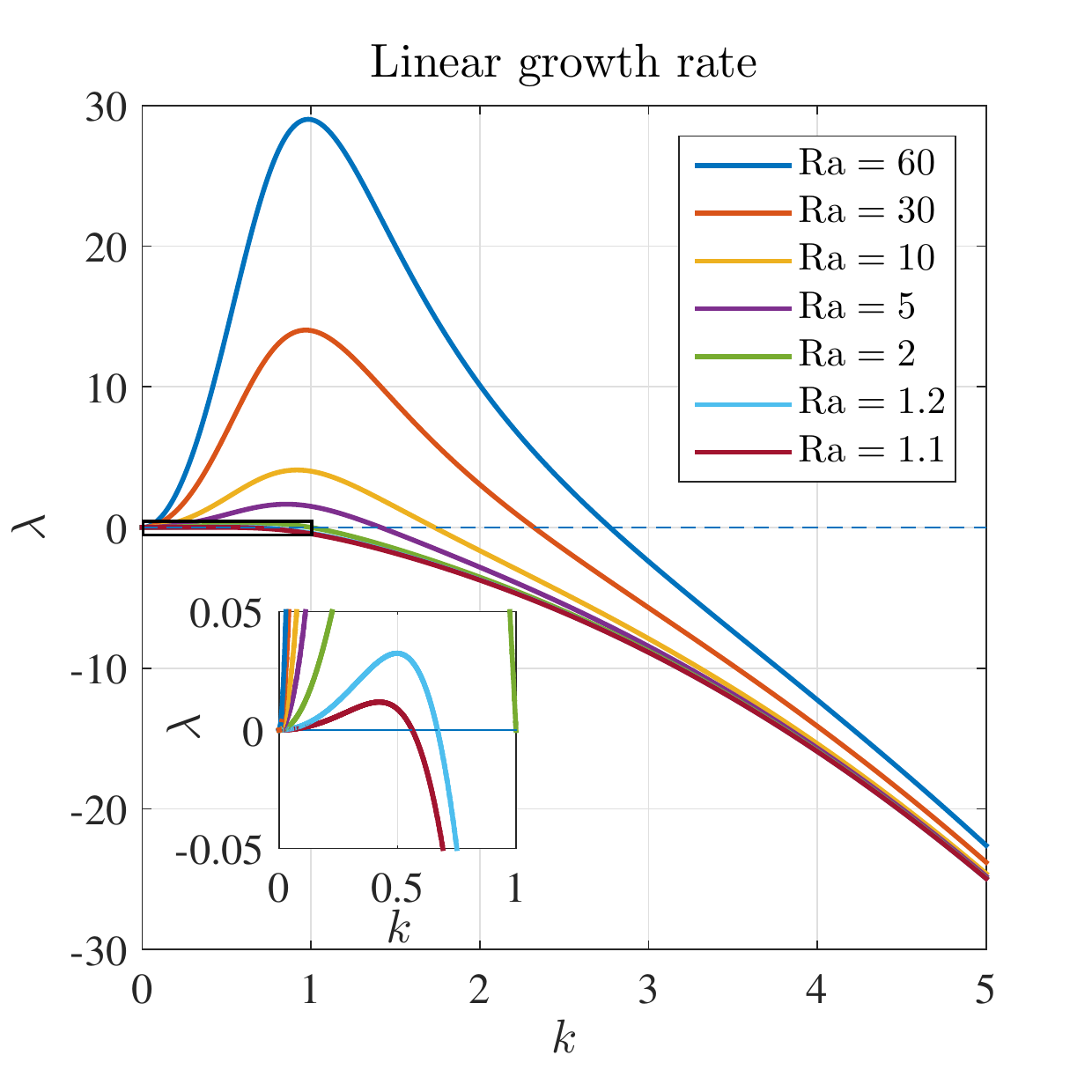}
	\includegraphics[width=0.4\linewidth]{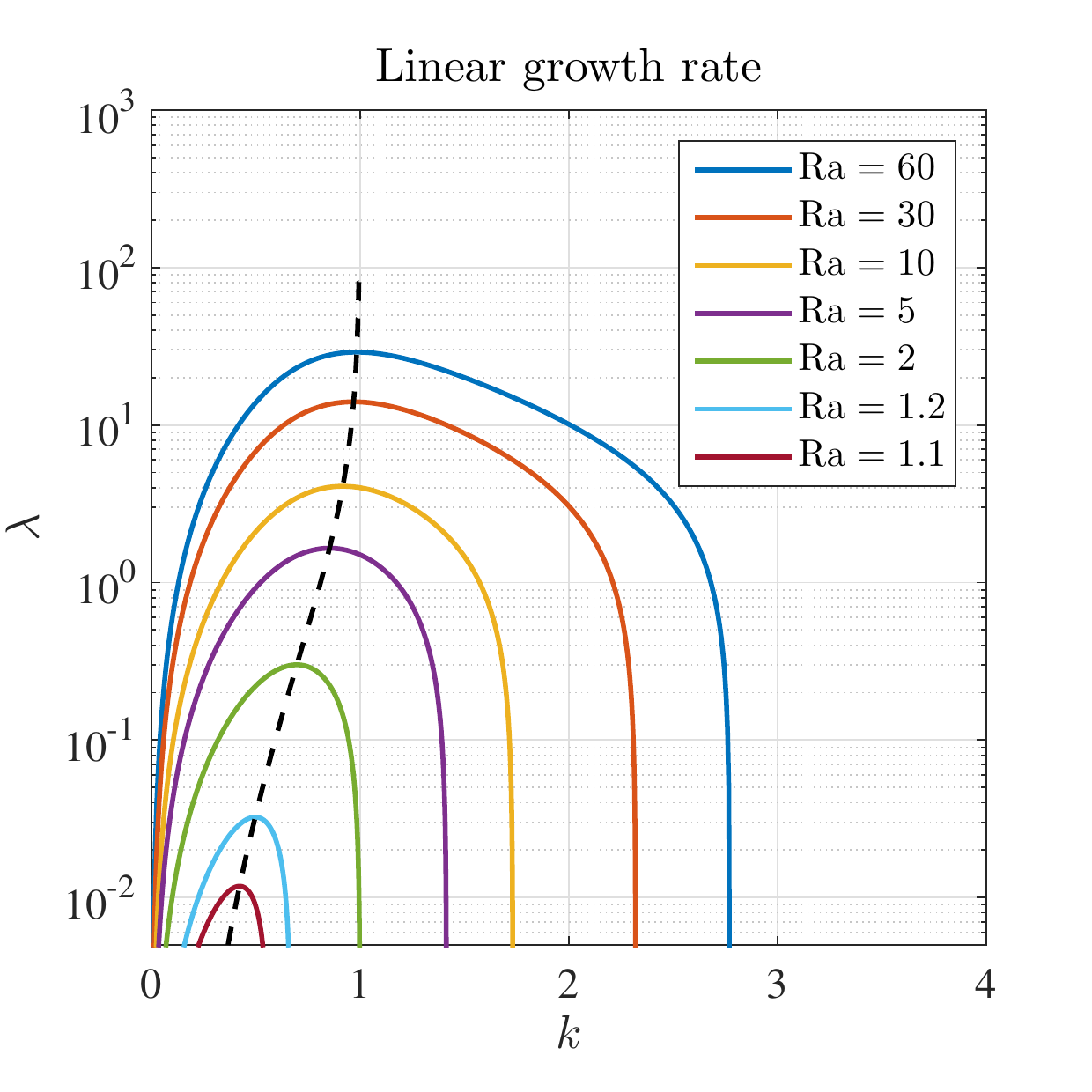}
	\caption{Linear growth rates $\lambda(k,0)$ of the elevator modes in the IFSC model (\ref{RedMod}) for different values of $\Ra$ in linear-linear plot (left) and linear-log plot (right). The inset in the left panel zooms in on the rectangular region where $\Ra$ is close to $1$; the black dashed line in the right panel indicates the fastest growing optimal mode.}
	\label{Linear_growth_Rate}
\end{figure}

We show the linear growth rate $\lambda(k,m)$ in Fig. \ref{Linear_Growth_Rate_2D} for (a) ${\Ra}=1.1$ and (b) ${\Ra}=10$.
Figure \ref{Linear_Growth_Rate_2D} highlights the anisotropic dependence of $\lambda$ on $k$ and $m$: for fixed $m$, not too large, there exists an optimal $k$ that maximizes the growth rate, whereas when $k$ is fixed, an optimal $m$ exists for small $k$ but for large $k$ the growth rate decreases monotonically with increasing $m$. For the elevator modes with zero vertical wavenumber, $m=0$, the band of unstable horizontal wavenumbers is well captured by the supercriticality $\R$:
\begin{equation}
0<k^4< \R. 
\end{equation}
As already mentioned, these modes are exact solutions of the nonlinear equations (\ref{RedMod}) with periodic boundary conditions in the vertical, and so play a potentially important role in the nonlinear regime. We present the growth rates of the elevator modes in Fig. \ref{Linear_growth_Rate} for several values of $\Ra$.

\begin{figure}
	\centering
	\includegraphics[width=0.4\textwidth]{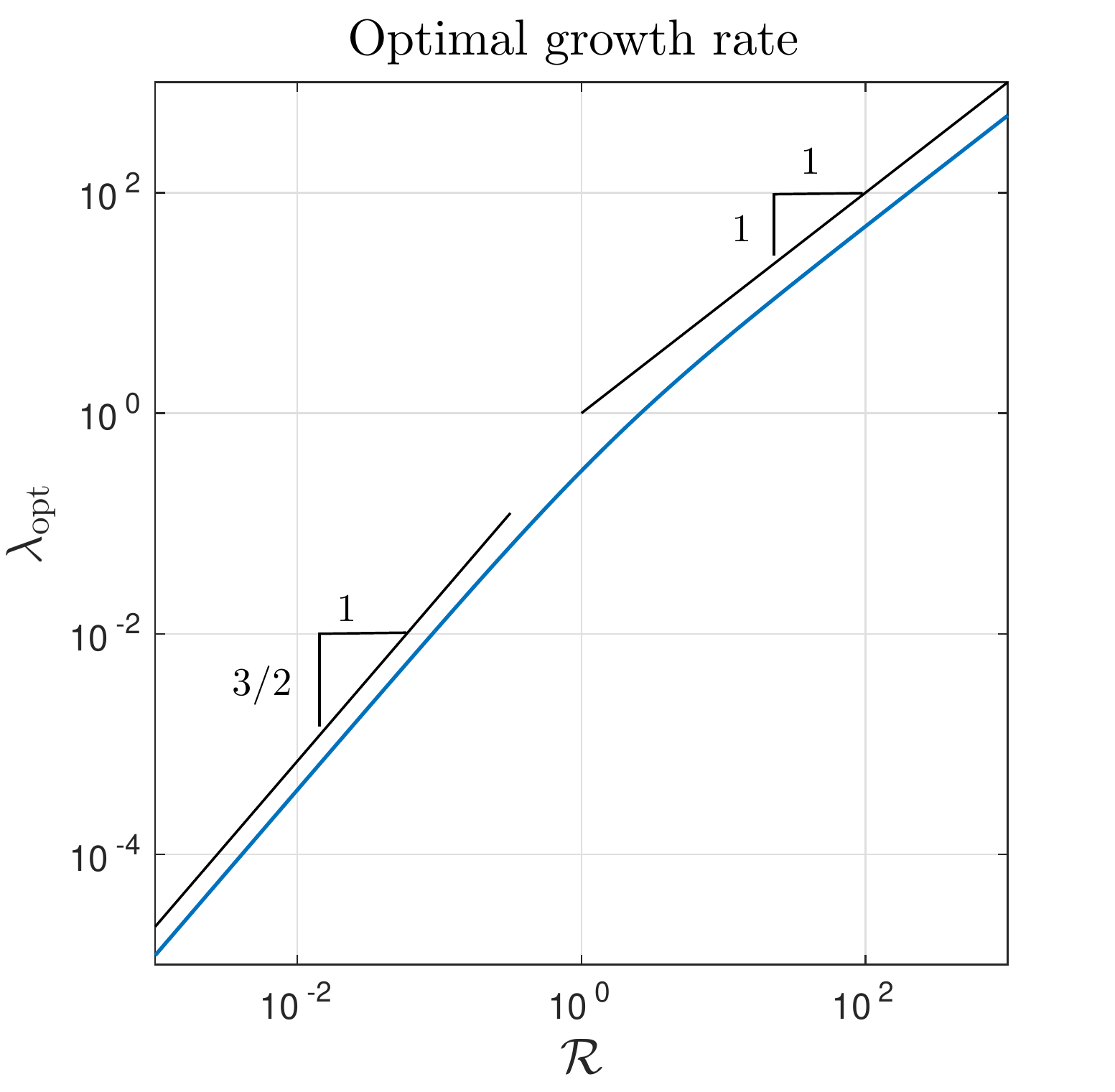}
	\includegraphics[width=0.4\textwidth]{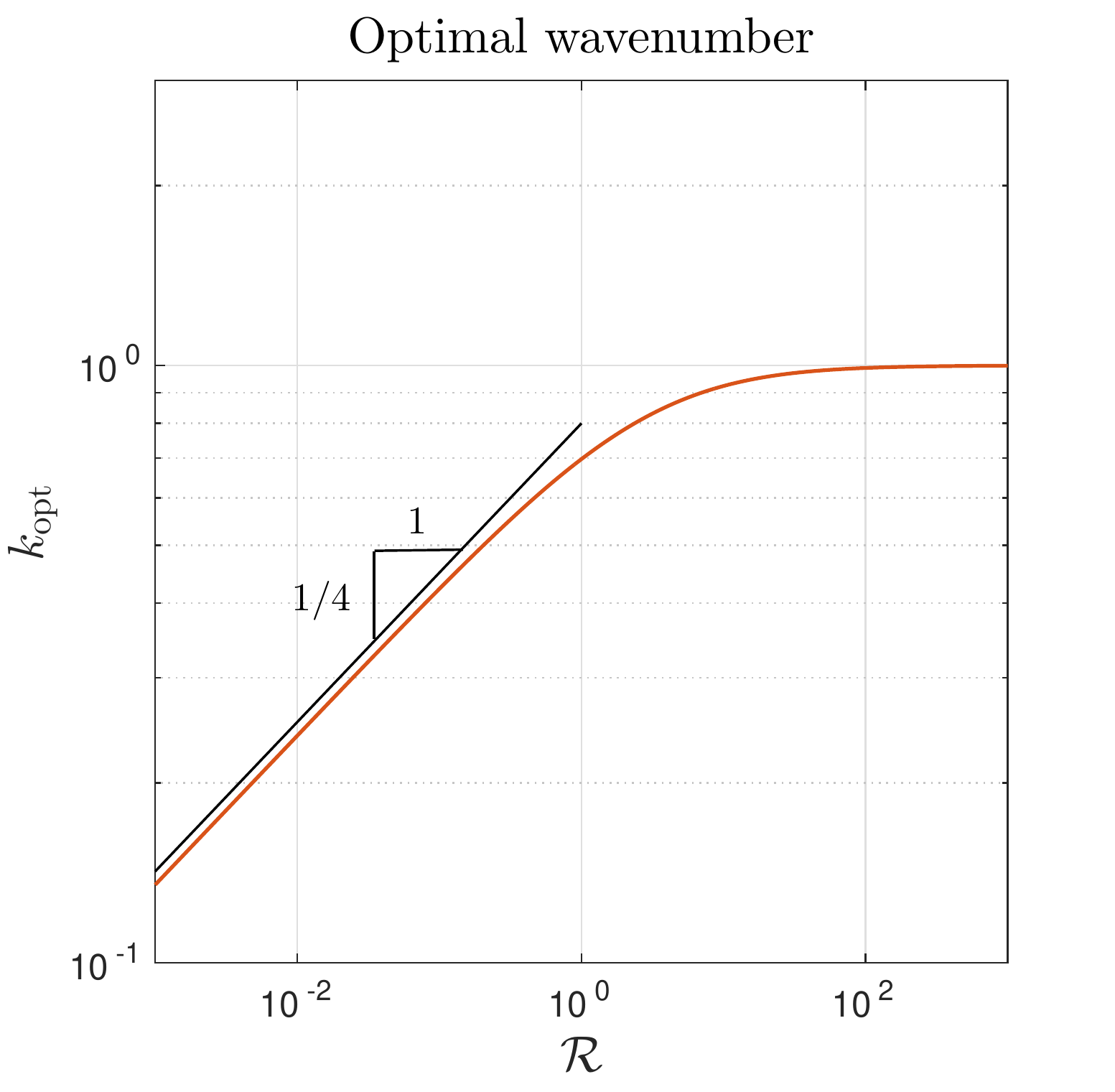}
	\caption{(a) Optimal growth rate $\lambda_\mathrm{opt}$ and (b) optimal wavenumber $k_\mathrm{opt}$ as functions of $\R\equiv\Ra-1$.}
	\label{Opt_lambda_k}
\end{figure}

In the absence of boundaries the fastest growing mode is an elevator mode such that
\begin{equation}
k_\mathrm{opt}^4 = \frac{1}{2}\br{-2-\Ra + \sqrt{\Ra^2 + 8\Ra}} \quad \mathrm{and} \quad m_\mathrm{opt}=0, \label{OptMode}
\end{equation}
with growth rate given by
\begin{equation}
\lambda_\mathrm{opt} = \sqrt{\frac{1}{2}\br{-2-\Ra + \sqrt{\Ra^2 + 8\Ra}}}\dfrac{3\Ra-\sqrt{\Ra^2 + 8\Ra}}{\sqrt{\Ra^2 + 8\Ra}-\Ra}\, . \label{Opt_growth_rate}
\end{equation}
In Fig. \ref{Opt_lambda_k} we replot these results in a log-log plot, indicating the asymptotic limits valid for $\R\ll1$ and $\R\to\infty$:
\begin{subequations}
	\begin{align}
	\lim\limits_{\mathcal{R}\rightarrow 0} k_\mathrm{opt} &= \frac{1}{3^{1/4}} \mathcal{R}^{1/4}
	, \qquad \lim\limits_{\mathrm{Ra}\rightarrow \infty} k_\mathrm{opt}  = 1
	\, ; \\
	\lim\limits_{\mathcal{R}\rightarrow 0} \lambda_{\mathrm{opt}} &= \frac{2}{3^{3/2}} \mathcal{R}^{3/2} 
	, \qquad 
	\lim\limits_{\mathrm{Ra}\rightarrow \infty}\lambda_{\mathrm{opt}} = \frac{\mathrm{Ra}}{2}.
	\end{align}
\end{subequations}

When stress-free, fixed temperature and fixed salinity boundaries conditions are applied on horizontal boundaries separated vertically by a gap $2\pi$, the marginal stability curve corresponds to $m_\mathrm{opt}=1$ and yields 
\begin{equation}
\Ra_\mathrm{mar} = 1 + \frac{\left(k^2 + 1\right)^3}{k^2}\, , \label{Mar_SF}
\end{equation}
a result highly reminiscent of the marginal stability curve of the classic B\'enard problem with similar boundary conditions: $\Ra^{(\mathrm{RBC})}_\mathrm{mar}=\left(k^2 + 1\right)^3/k^2$ \cite{Chandrasekhar1961}.  

We conclude this section with an important energy argument. In a doubly periodic domain with period $L_x$ in the horizontal direction and $L_z$ in the vertical direction, we define the spatially averaged salinity potential energy $E_S$ as:
\begin{equation}
E_S = \frac{1}{L_xL_z} \int  \frac{1}{2}S^2(x,z,t) \dd x \dd z.
\end{equation}
For $\boldsymbol{k}\in \left(2\pi n_x/L_x, 2\pi n_z/L_z \right)$ with integers $n_x$ and $n_z$, one defines the Fourier series $\hat{S}_{\boldsymbol{k}}$:
\begin{subequations}
	\begin{gather}
	\hat{S}_{\boldsymbol{k}}(t) = \frac{1}{L_x L_z} \int S(x,z,t)\mathrm{e}^{\mathrm{i\boldsymbol{k \cdot r}}} \dd^2 \boldsymbol{r}, \\
	S(x,z,t) = \sum_{\boldsymbol{k}} \hat{S}_{\boldsymbol{k}}(t)\mathrm{e}^{-\mathrm{i\boldsymbol{k \cdot r}}}\, ,
	\end{gather}
\end{subequations} 
where $\boldsymbol{r}=(x,z)$,
such that the average energy can be expressed as the sum of contributions in spectral space, as a consequence of Parseval's theorem:
\begin{equation}
E_S = \frac{1}{2}\sum_{\boldsymbol{k}} \left| \hat{S}_{\boldsymbol{k}}\right| ^2.
\end{equation}
Thus the energy in wavenumber $\bk$ is $\hat{E}_S(\bk) = \frac{1}{2}\left| \hat{S}_{\boldsymbol{k}}\right| ^2$.

A spectral energy budget is readily obtained by taking a Fourier transform of Eq.~(\ref{Pro_Rel}), multiplying each Fourier component by the complex conjugate $\hat{S}^*_{\bk}$ and summing over $\bk$, yielding:
\begin{equation}
\frac{\dd }{\dd t} E_S =\sum_{\boldsymbol{k}} \lambda(\bk) \left|{\hat{S}}_{\boldsymbol{k}}\right|^2 \, , 
\end{equation}
where $\lambda$ is the growth rate (\ref{GroRat}). This energy equation implies that in the saturated state, the energy input from unstable modes compensates energy dissipation by damped modes, with the advection term transferring energy between them.

We define the spatially-averaged vertical salinity flux $F_S$ as
\begin{equation}
F_S = \frac{1}{L_xL_z} \int  \psi_xS \dd x \dd z.
\end{equation}
This flux can be expressed as
\begin{equation}
F_S = \sum_{\boldsymbol{k}} \hat{F}_S(\bk),
\end{equation}
where
\begin{equation}
\hat{F}_S(\bk) = \frac{\ii k}{2}\br{ \hat{\psi}_{\bk}\hat{S}^*_{\bk} - \hat{\psi}^*_{\bk}\hat{S}_{\bk} } = -\frac{k^2\abs{K}^2}{k^2+\abs{K}^6}\abs{\hat{S}_{\bk}}^2,
\end{equation}
using the diagnostic relation (\ref{Diagnostic_balance}). Thus $F_S<0$, as expected.

\subsection{Secondary instability}

\begin{figure}
	\centering
	\includegraphics[width=\linewidth]{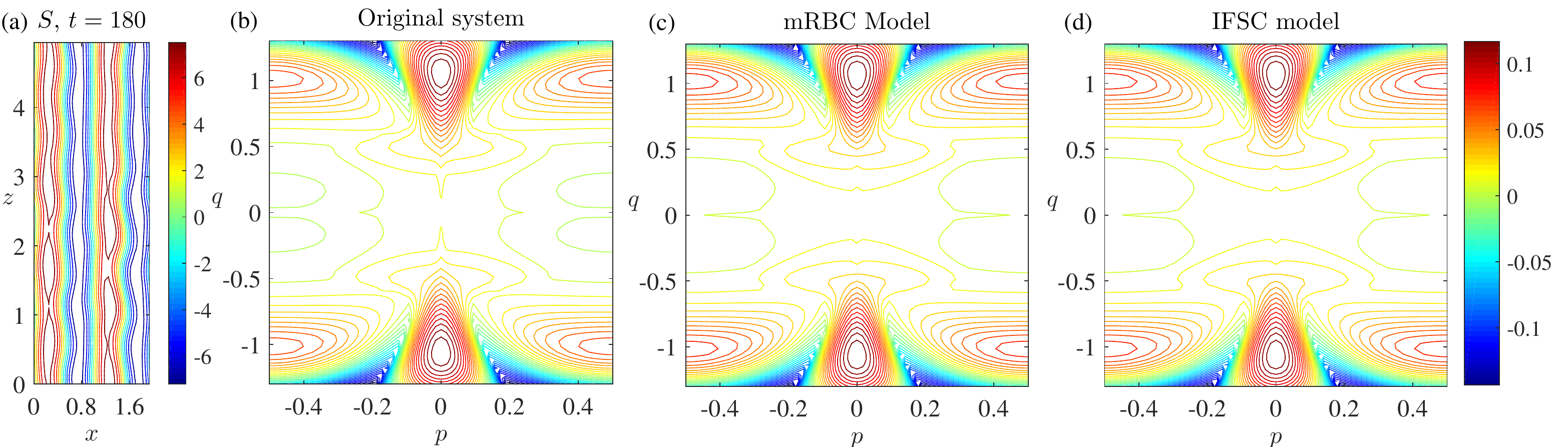}
	\caption{(a) The salinity field in the early stages ($t=180$) of the development of a secondary instability in a $2\ell_\mathrm{opt}\times 5\ell_\mathrm{opt}$ domain when $\Ra=1.1$. Contours of constant growth rate $\alpha$ in the $(p,q)$ plane for (b) the primitive system (\ref{GovEq}), (c) the modified RBC equations (\ref{mRBC}) and (d) the IFSC model (\ref{RedMod}). Wavenumbers are normalized by the optimal wavenumber $k_\mathrm{opt}$ given by (\ref{OptMode}).}
	\label{2ndIns}
\end{figure}

At some point the growing elevator modes will trigger secondary instabilities that may be responsible for the break up of the exponentially growing elevator modes and the saturation of the instability. Figure ~\ref{2ndIns}(a) shows the salinity field obtained from a simulation of the IFSC model (\ref{RedMod}) in a $2\ell_\mathrm{opt}\times 5\ell_\mathrm{opt}$ domain with an initial condition that is a combination of the optimal mode and superposed small amplitude random noise. Here $\ell_\mathrm{opt}\equiv2\pi/k_\mathrm{opt}$. The figure shows that at $t=180$ a secondary instability sets in with a vertical wavenumber comparable to that of the optimal mode. As an indication of what secondary instabilities may be present we assume a quasi-static approximation where the salt fingers are considered steady. This is a reasonable approximation in the sense that when the secondary instability is important its growth rate should be larger than that of the salt fingers. We analyze the stability of this state within both the IFSC model (\ref{RedMod}) and the modified RBC system (\ref{mRBC}) by means of single vertical mode Floquet theory \cite{Holy1984}, which considers perturbations of the form 
\begin{equation}
\begin{pmatrix}
\psi \\
S
\end{pmatrix}
=
\ex^{\alpha t + \ii k\br{px+qz}} \sum_{n=-N}^{n=N}
\begin{pmatrix}
\psi_n \\
\ii S_n
\end{pmatrix}
\ex^{\ii n kx}.
\label{FloqExpan}
\end{equation}
Comparisons are made with the full system~(\ref{GovEq}) with $\tau = 0.01$,  $\mathrm{Ra} = 1.1$, and $\mathrm{Sc}=100$ which corresponds to $\mathrm{Pr}=1$ and $R_\rho = 0.011^{-1}\approx 90.91$. Given that $\tau$ is an asymptotically small parameter, the corresponding mRBC model requires $\mathrm{Ra} = 1.1$ and $\mathrm{Sc}=100$ while the corresponding IFSC model only requires $\mathrm{Ra}=1.1$.

We observe in time-stepping simulations of the IFSC model that the secondary instability of the elevator modes is triggered at a salinity amplitude of approximately $S_e \approx 6.5$. This amplitude is selected as the common amplitude for the Floquet theory in all models. For the primitive equations, this implies a renormalization by a factor of $\mathrm{Ra}$ (see Eq.~(\ref{T_S_renormalized})).

In Fig. \ref{2ndIns}(b)--(d), we show a contour plot of the growth rate $\alpha$ of the secondary instability as a function of the Floquet wavenumbers $p$ and $q$. The observed match between the three panels demonstrates the validity of the reduced models. Moreover, the prediction of the single vertical mode Floquet theory with $S_\mathrm{e}=6.5$ that the largest growth rate occurs for $p=0$ and $q=1.05$ is in excellent agreement with the onset of the observed secondary instability. For comparison we mention that for $\Pr=7$, $\Rr=2$ and $\tau=1/24$ \citet{Ster2005} found that the secondary instability sets in with vertical wavenumber $q=0.8$, expressed in units of the optimal wavenumber. These results confirm the validity of the single vertical mode Floquet theory in describing the onset of the secondary instability.

\subsection{Towards saturation} \label{SecTowSat}

The saturated states of the IFSC model (\ref{RedMod}) depend on the domain size: when the domain is small a steady state is reached; in an intermediate size domain chaotic states are observed; for sufficiently large domains the system reaches statistically steady states involving a broad range of scales. The results reported below are carried out in domains of sufficiently large size that the results are independent of the domain size used. For comparison we summarize the corresponding small-domain and intermediate-domain results in Appendix \ref{Sec_domain_size}. 

In Fig.~\ref{Energy_field} we show the results obtained for $\Ra=1.1$ in a domain of size $32\times32$ in units of the optimal wavelength $\ell_\mathrm{opt} = 14.86$ with an initial condition in the form of a small amplitude random field. We ran the simulation to $t=4800$, which is large compared to the characteristic time $\lambda_\mathrm{opt}^{-1}=84.84$ for the growth of the optimal mode, and observed (Fig. \ref{Energy_field}(a)) that a statistically steady state is finally reached. Three stages toward saturation can be identified: 
(i) {\it Dominance of salt fingers}, $t\lesssim 138$. This is the stage corresponding to peak energy and flux generation. Figure \ref{Energy_field}(b) shows the salinity field in the vicinity of this peak $(t\approx 138)$; long finger structures are observed. 
(ii) {\it Secondary instabilities}, $138 \lesssim t\lesssim 500$. The salt fingers cannot grow without bound owing to the onset of secondary instability, whose growth rate increases with the amplitude of the growing fingers as discussed in detail in the previous section. 
(iii) {\it Saturated state}, $t\gtrsim 500$. This regime is characterized by collisions of upward and downward fingers \cite{Shen1995}.
Figure \ref{Energy_field}(c) provides a snapshot of the final statistically steady salinity field. Finger-like structures can still be recognized but these are of small vertical extent and quite different from those present in stage (i). It appears that it is a combination of these two effects -- secondary instability and salt finger collision -- that is responsible for the observed statistically steady saturated state. In an alternative argument \citet{Radk2003} suggests that saturation occurs when the growth rates of the linear fingering instability and the secondary instability become comparable. Such a balance implies a balance between linear and nonlinear terms on the same horizontal scale, but as discussed in \S \ref{Sec_Dis} this balance is not always satisfied.
\begin{figure}
	\centering
	\includegraphics[width=1\textwidth]{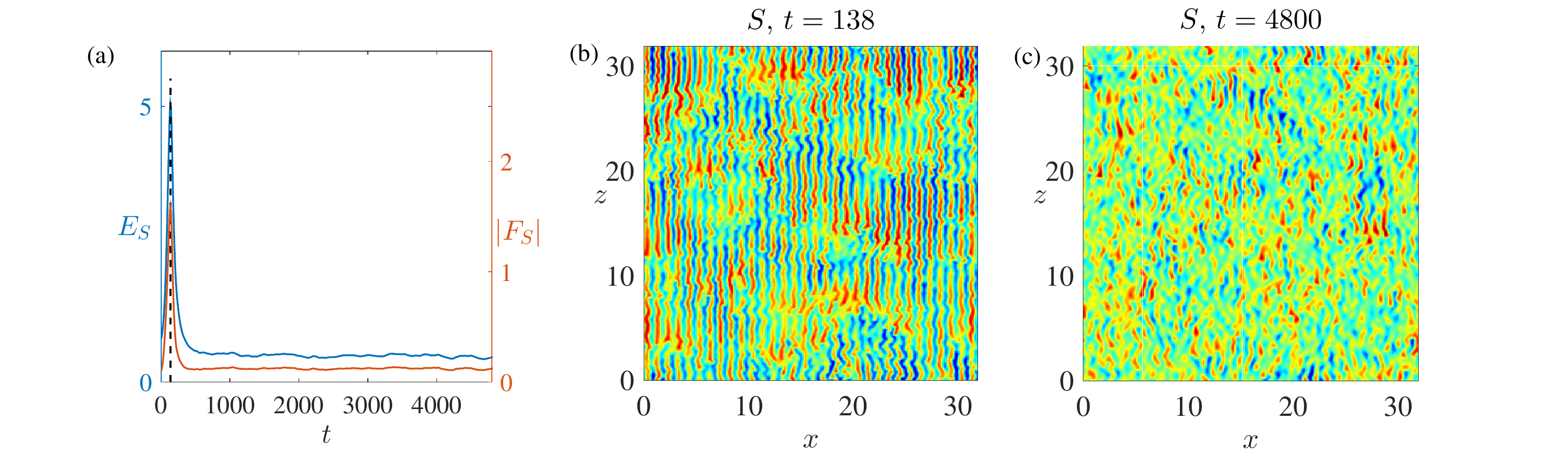}
	\caption{Results from a $32\ell_\mathrm{opt}\times32\ell_\mathrm{opt}$ simulation for $\Ra=1.1$. (a) Evolution of the total energy $E_S$ and the salinity flux $|F_S|$ with time. The dashed line marks $t=138$, where both quantities peak. (b) The finger-dominated state at $t=138$. (c) The instantaneous statistically steady state at $t=4800$, the end of this simulation. Distances are measured in units of $\ell_\mathrm{opt}$, the wavelength of the optimal mode.}
	\label{Energy_field}
\end{figure}

In the next section we characterize the large-domain saturated state in more detail.

\section{Saturated states in the inertia-free salt convection model} \label{Sec_SatSta}

In large domains (i.e., for large thermal Rayleigh numbers $\Ra_T$, cf. Eq.~(\ref{RaT})), the IFSC model (\ref{RedMod}) reaches domain size-independent statistically steady states, whose statistical properties depend on the Rayleigh ratio $\Ra$. In this section we study this $\Ra$ dependence, since this is crucial for a sound parameterization of small-scale salt-fingering convection and its effect on large-scale fields. Global quantities -- available potential energy and flux -- are studied in \S \ref{Sec_Regimes}, where we identify two distinct regimes according to their dependence on $\Ra$. The corresponding probability density functions are shown in \S \ref{Sec_pdf}.

\begin{figure}
	\begin{center}
		\includegraphics[width=0.32\textwidth]{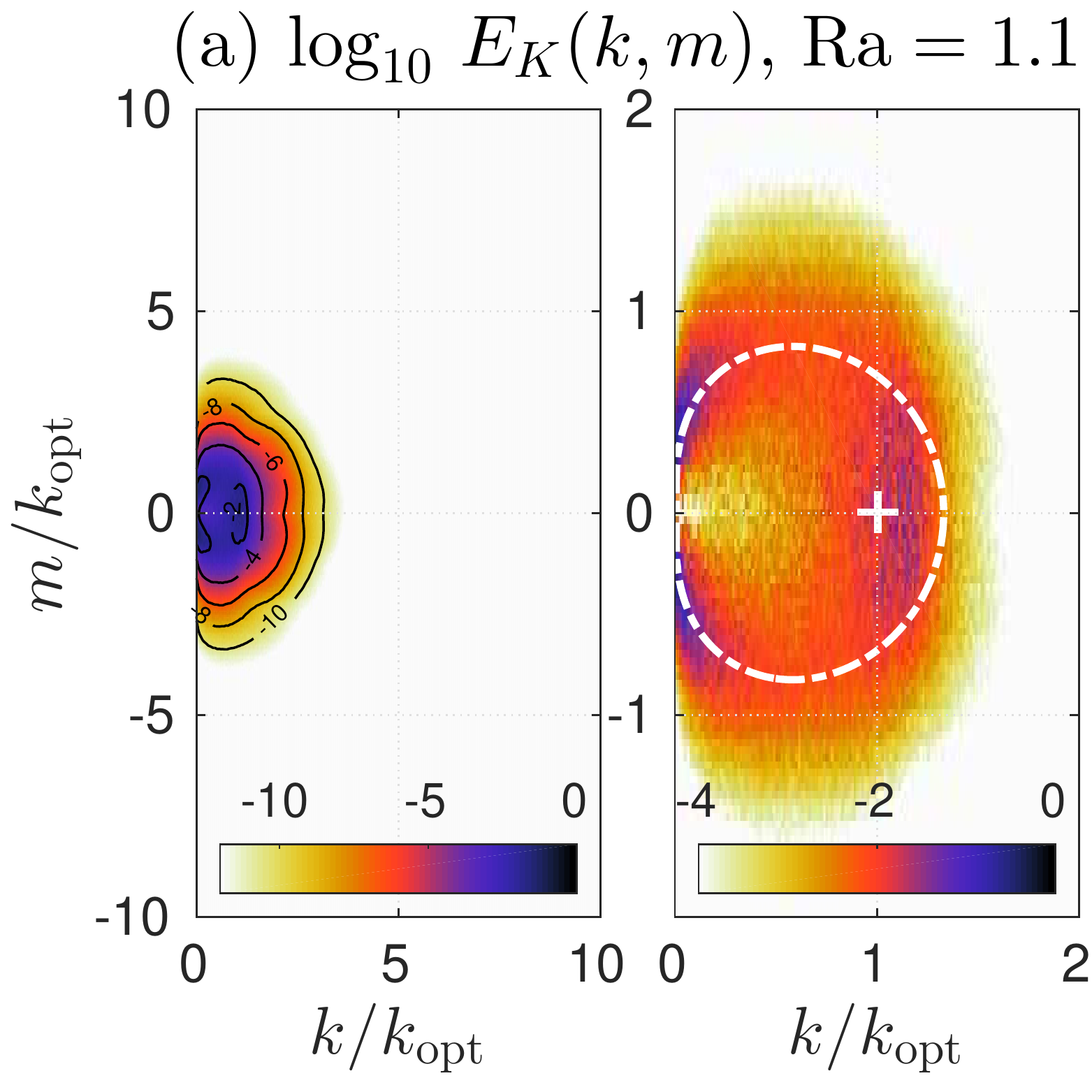}
		\includegraphics[width=0.32\textwidth]{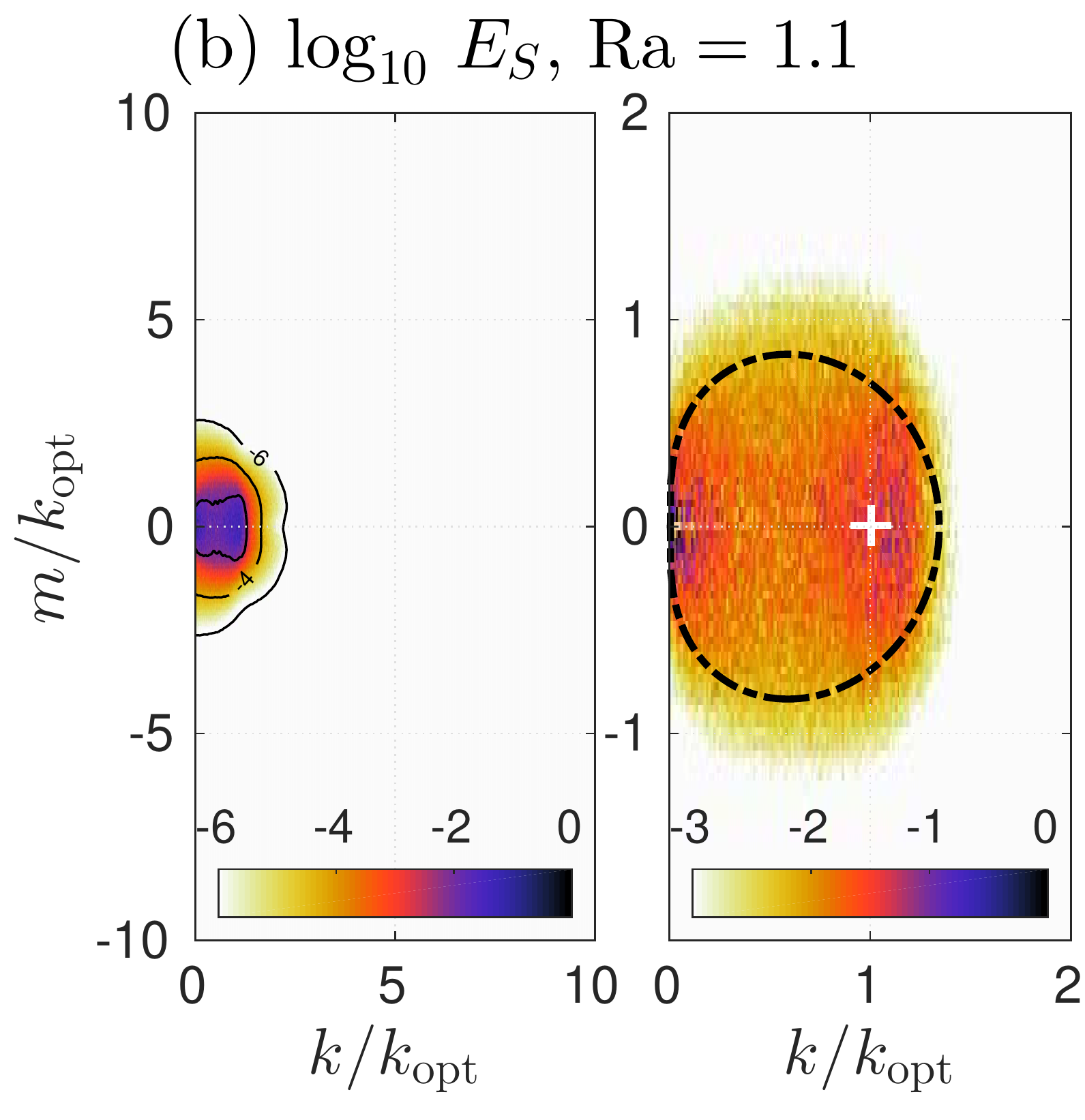}
		\includegraphics[width=0.32\textwidth]{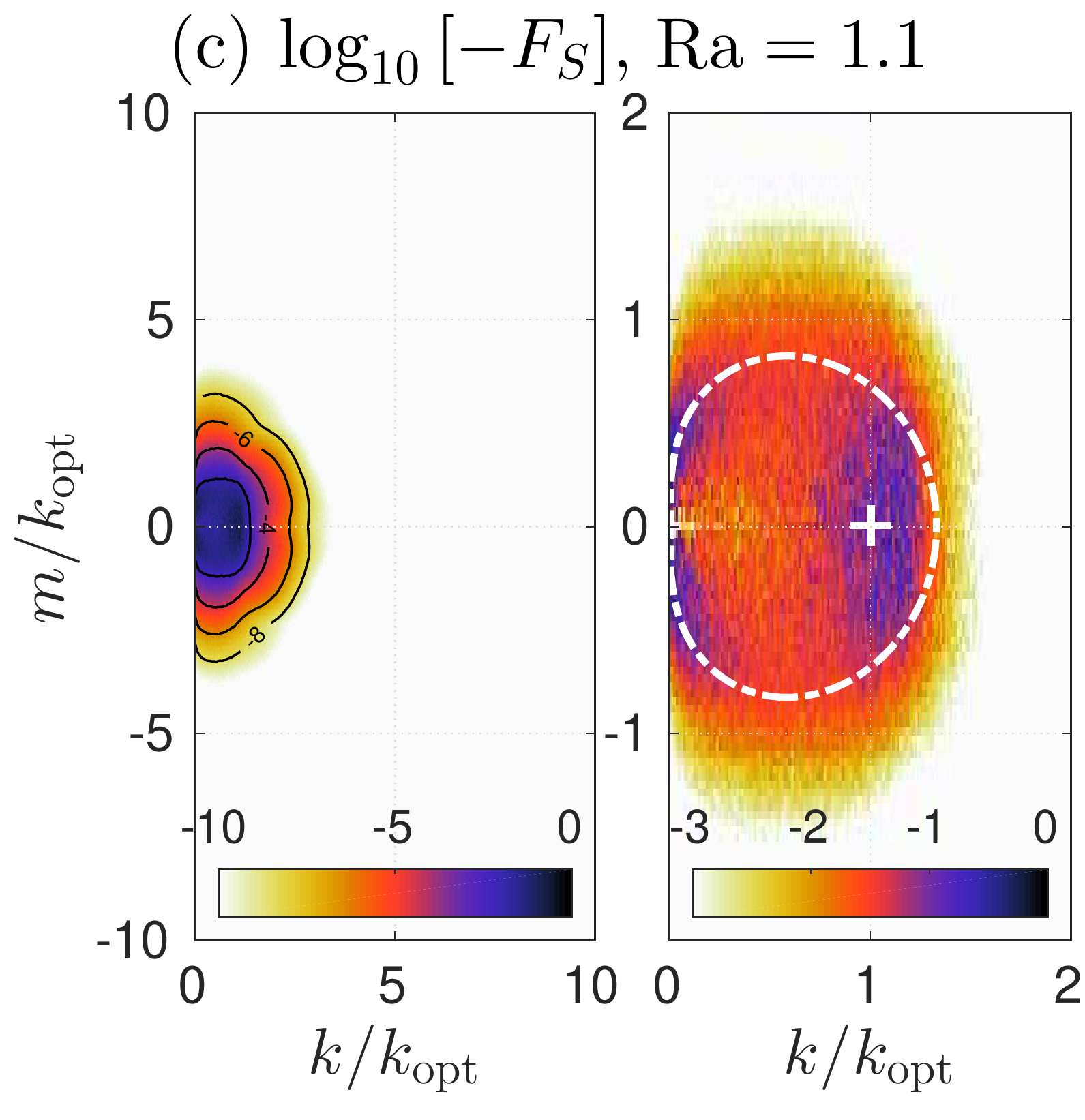}
		\includegraphics[width=0.32\textwidth]{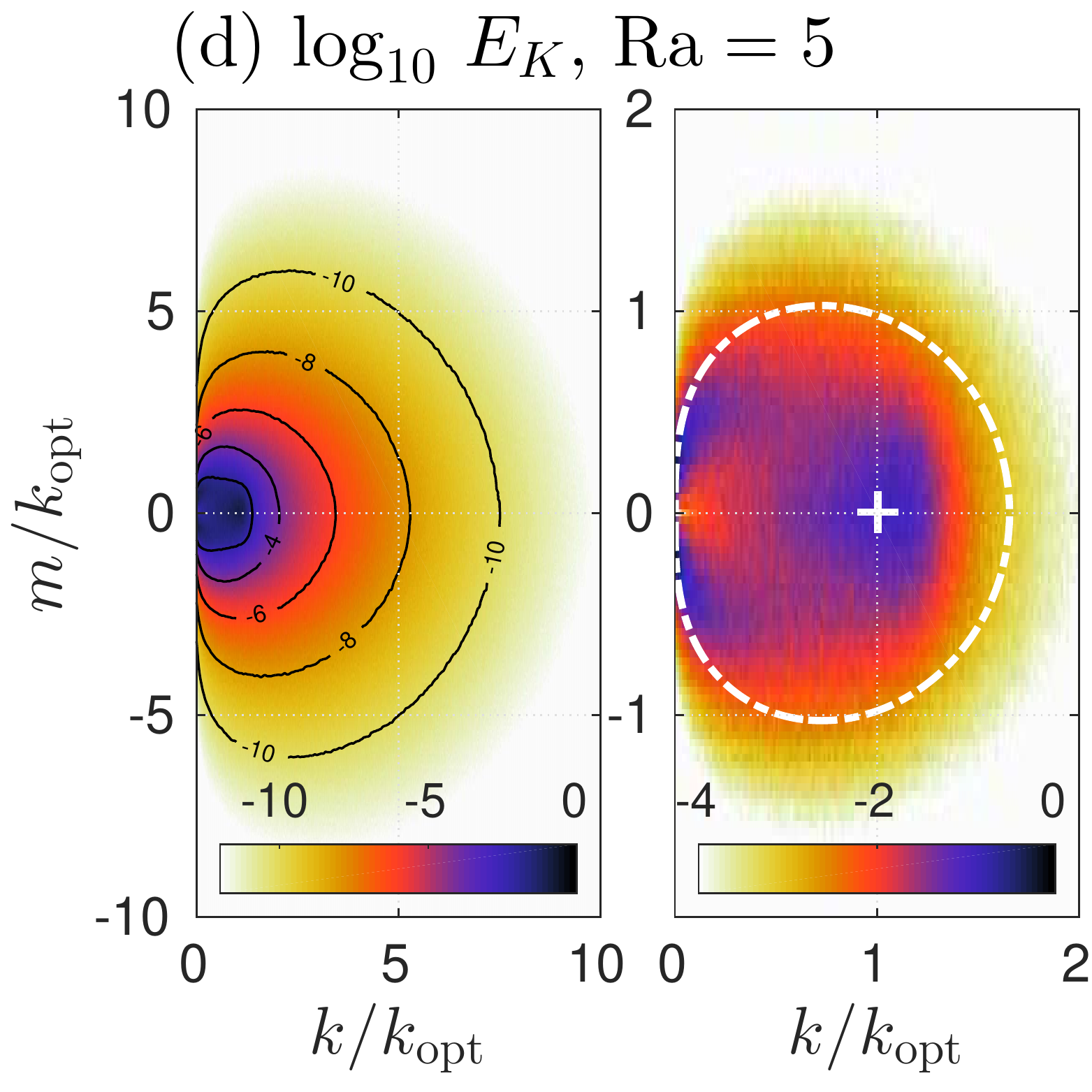}
		\includegraphics[width=0.32\textwidth]{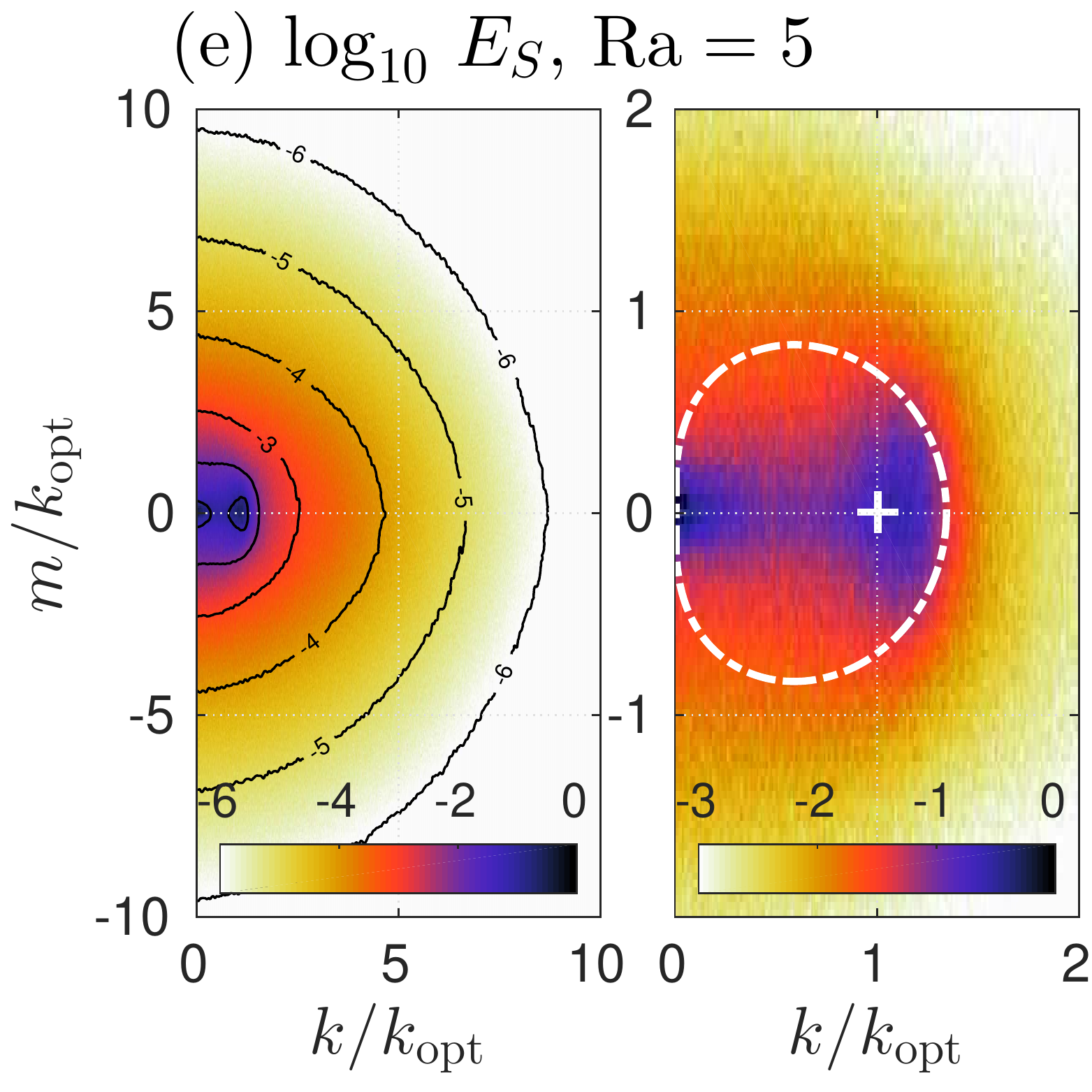}
		\includegraphics[width=0.32\textwidth]{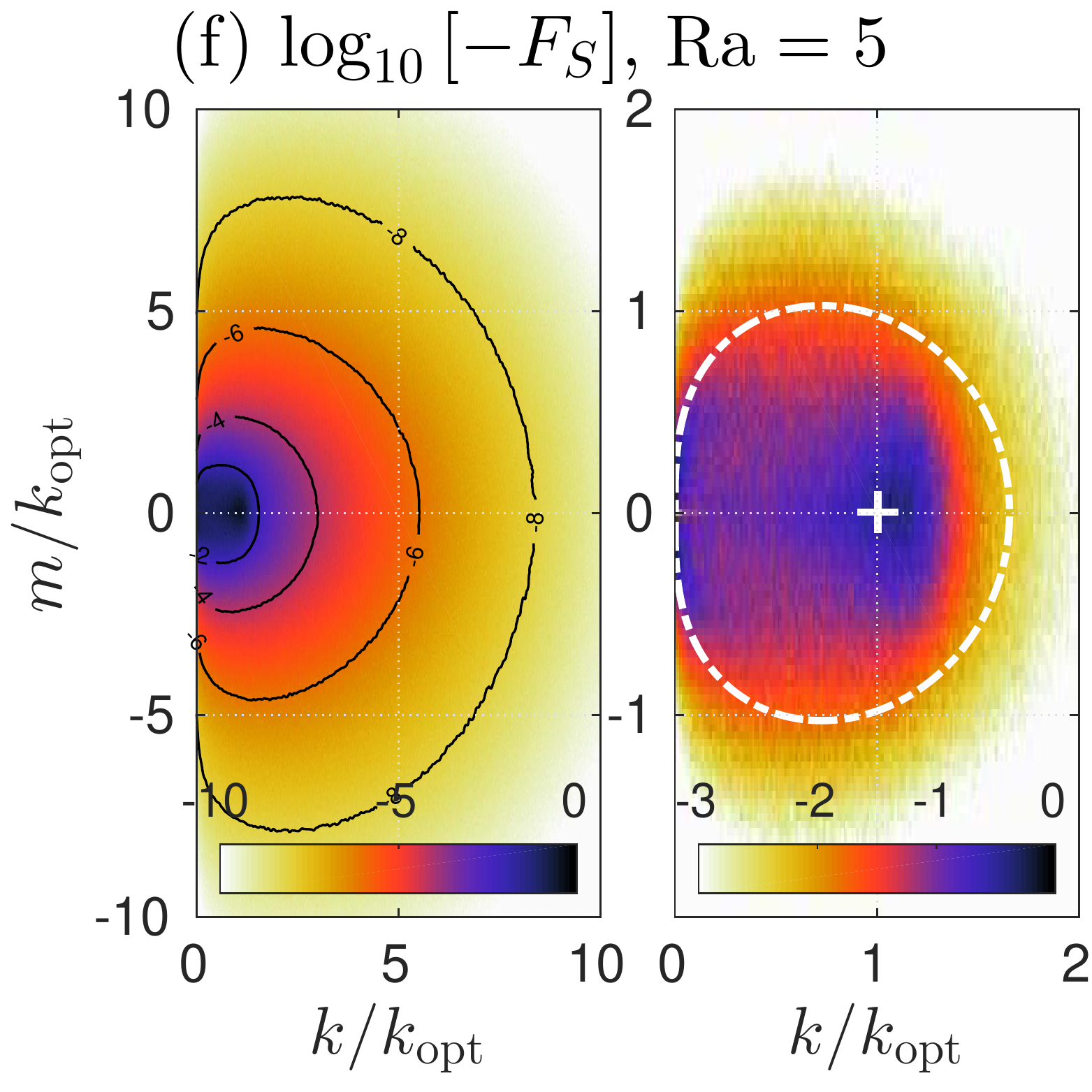}
		\includegraphics[width=0.32\textwidth]{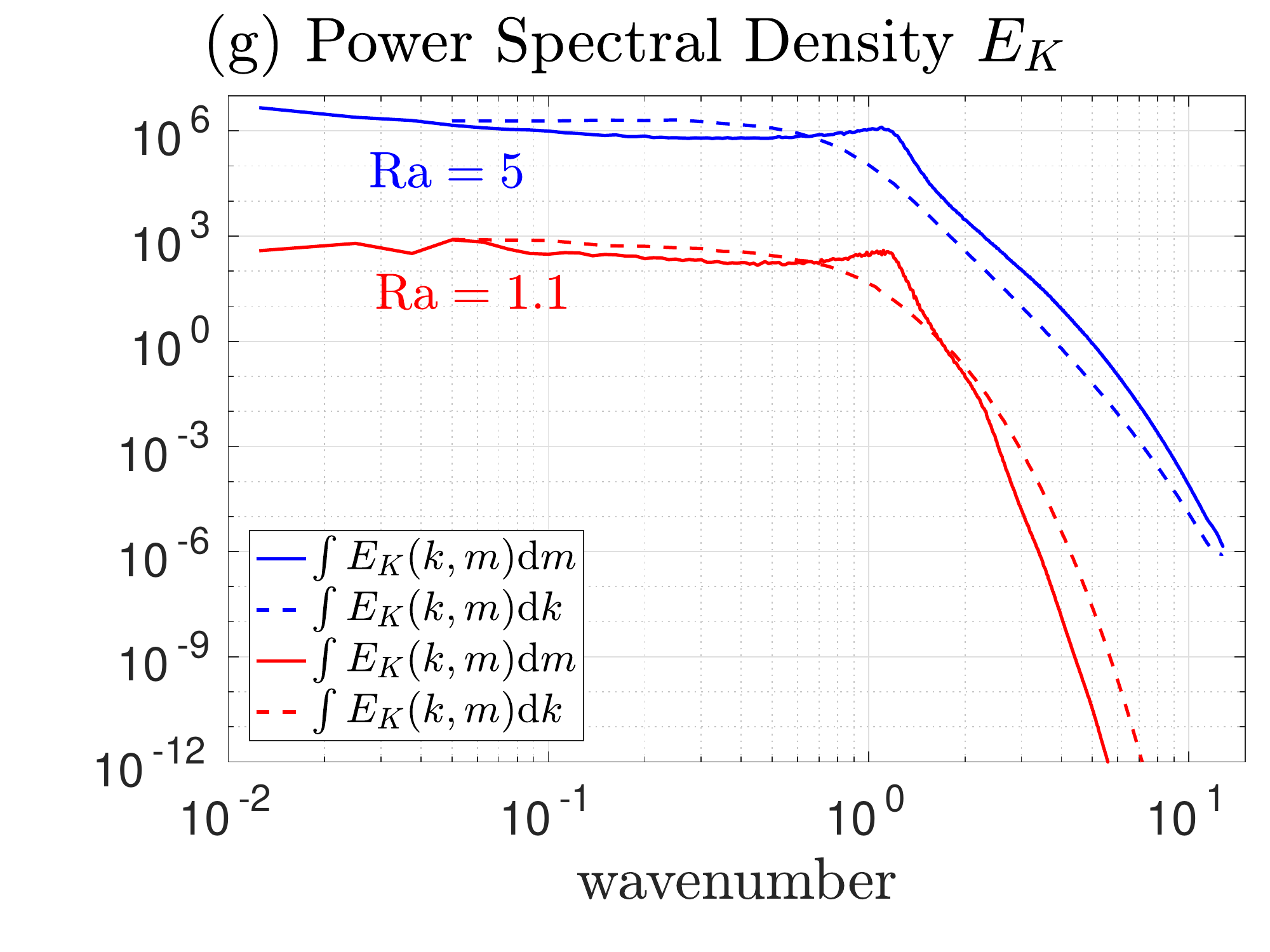}
		\includegraphics[width=0.32\textwidth]{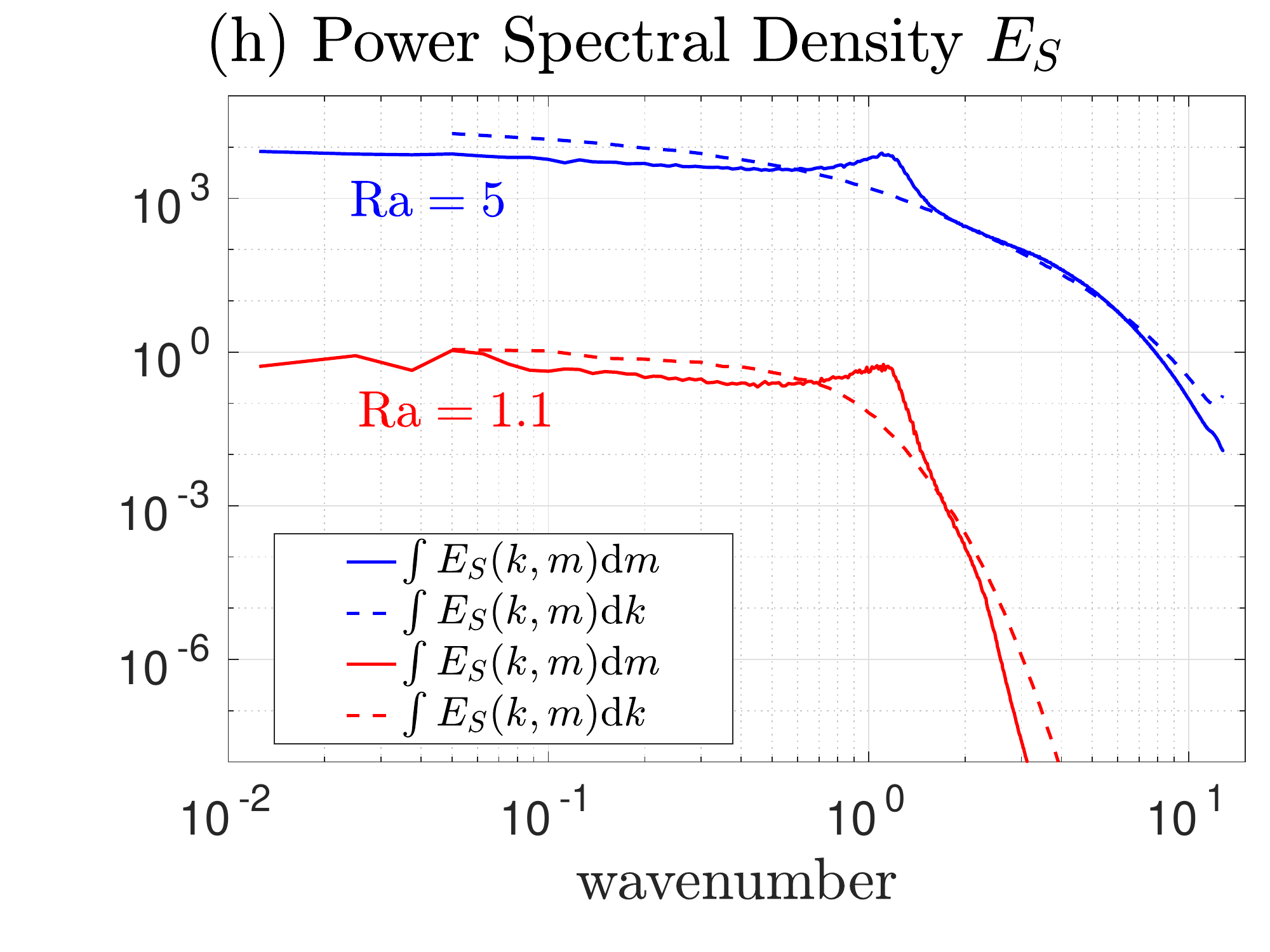}
		\includegraphics[width=0.32\textwidth]{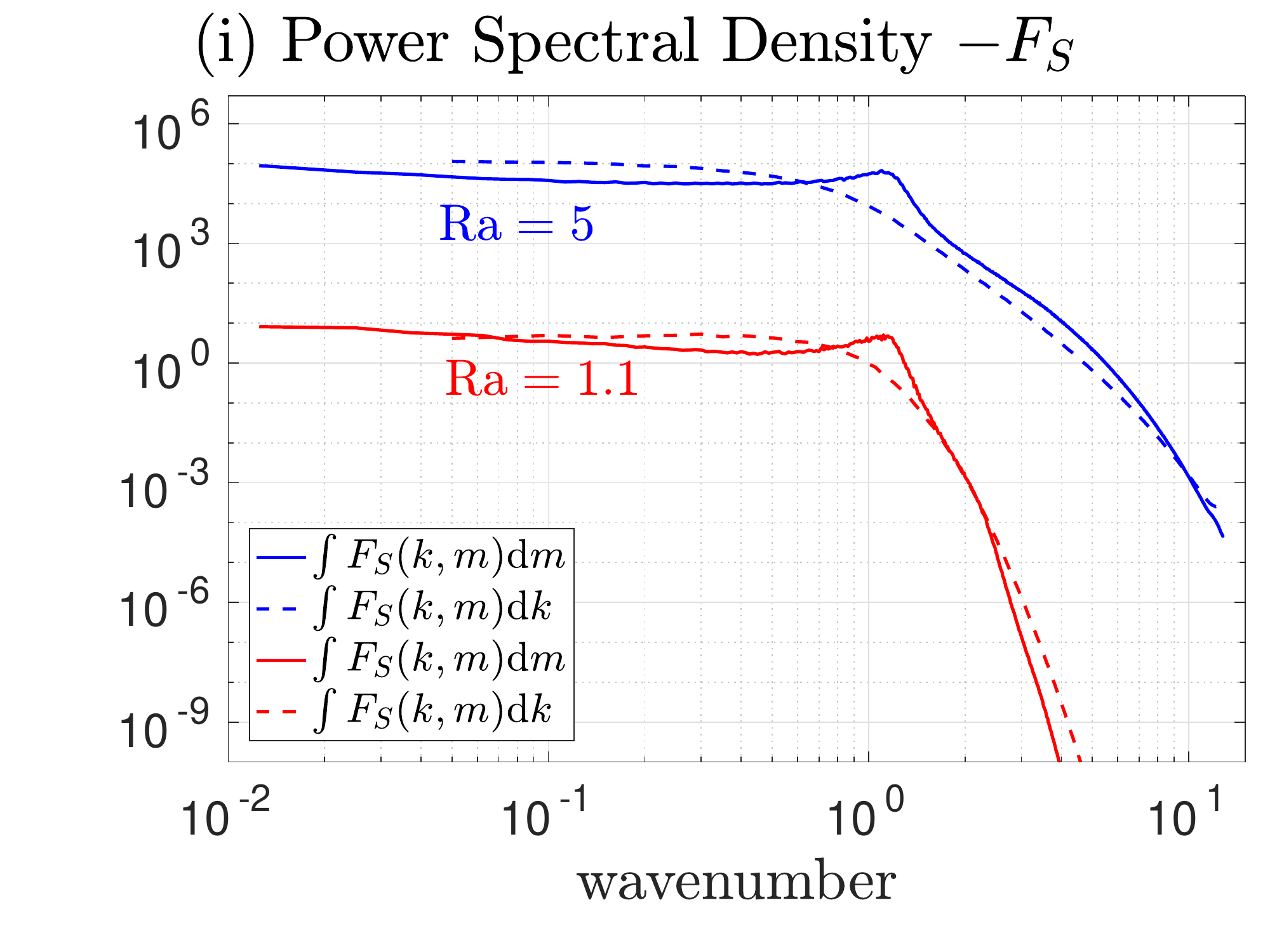}
	\end{center}
	\caption{From left column to right column: Spectral density of kinetic energy (a, d and g), salinity potential energy (b, e and h) and salinity flux (c, f, and i). The top row  and middle row correspond to $\mathrm{Ra}=1.1$ and $\mathrm{Ra}=5$, respectively, and are organised as follows: each subfigure consists of two panels, the left of which contains a bidimensional spectrum of one of the aforementioned three quantities while the right panel contains a close-up of the left panel. In the close-up panels, the thick dashed lines correspond to the marginal stability boundary, and the cross indicates the most unstable mode. Bottom row: corresponding integrated spectra along the vertical (plain lines) and horizontal (dashed lines) directions, plotted as a function of the vertical and horizontal wavenumber, respectively, for $\mathrm{Ra}=1.1$  (red lines) and $\mathrm{Ra}=5$ (blue lines).}
	\label{Spectrum}
\end{figure}

\begin{figure}
	\begin{center}
		\includegraphics[width=0.64\textwidth]{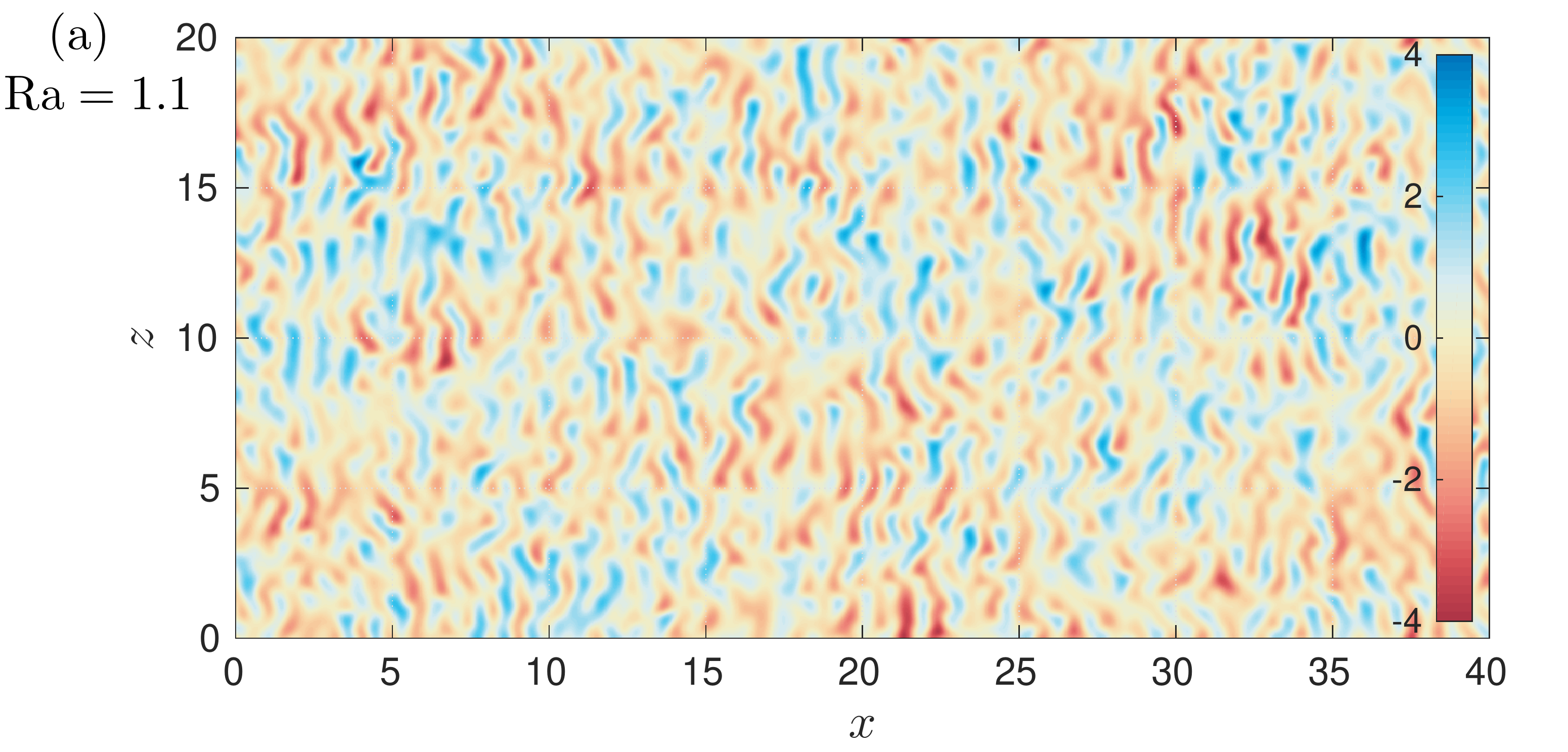}
		\includegraphics[width=0.64\textwidth]{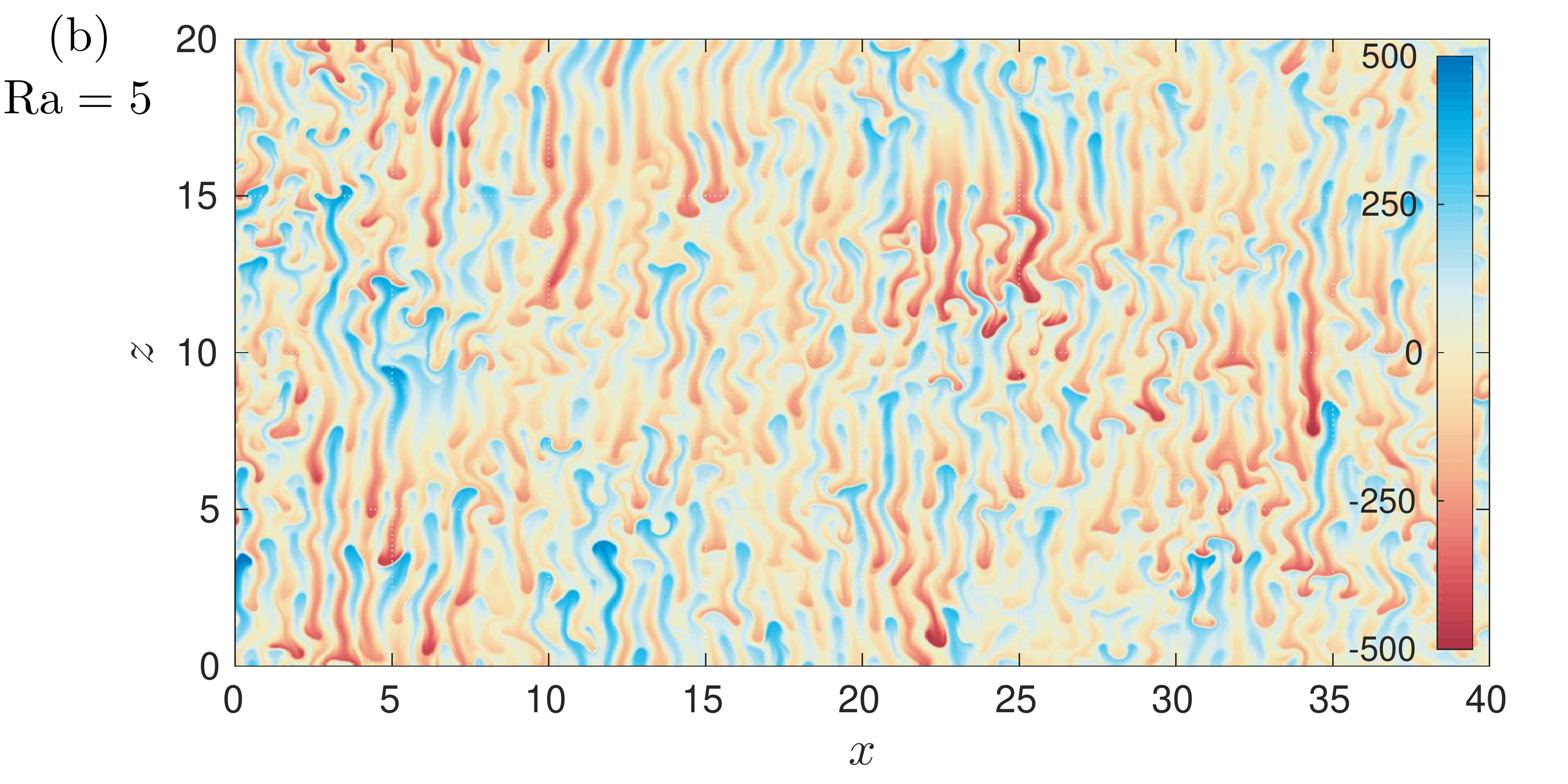}
	\end{center}
	\caption{Snapshot of the salinity perturbation for $\mathrm{Ra}=1.1$ (top panel) and $\mathrm{Ra}=5$ (bottom panel).}
	\label{Field}
\end{figure}

Figure \ref{Spectrum} shows the spectral density of kinetic energy (first column), salinity potential energy (second column) and salinity flux (third column) in the statistically stationary state as a function of the horizontal wavenumber $k$ and the vertical wavenumber $m$ (top two rows).  The third row shows the corresponding integrated spectra along the vertical (plain lines) and horizontal (dashed lines) directions, plotted as a function of the vertical and horizontal wavenumber, respectively. One distinguishing feature of the vertically integrated spectra is that they peak around the optimal wavenumber $k_\mathrm{opt}$. Away from the peak, the vertically integrated spectra increase weakly for wavenumbers below $k_{\rm opt}$, but fall off rapidly for wavenumbers larger than this energy injection wavenumber. Similar tendencies are also observed in the horizontally integrated spectra. At smaller Ra ($\mathrm{Ra}=1.1$) this fall-off is more rapid with increasing horizontal wavenumber than with increasing vertical wavenumber, a situation that reverses for larger Ra ($\mathrm{Ra}=5$). Of particular interest is the fact that energy is transferred to ever larger scales and in fact appears to peak at these scales. We surmise that this is a consequence of the patchiness of the saturated state (Fig.~\ref{Energy_field}(c)), a property that is also observed in three-dimensional simulations of the primitive equations \cite{Paparella2012}. Figure \ref{Field} compares the salinity fields in the stationary state for $\mathrm{Ra}=1.1$ (Fig.~\ref{Field}(a)) and $\mathrm{Ra}=5$ (Fig.~\ref{Field}(b)). The fingers are clearly visible, as are collisions between descending salt-fingers and rising fresh-water fingers. Large-scale patchy structures with fingers of finite length are also observed, indicating the presence of multiscale dynamics that is characteristic of Regime II, cf. \S \ref{SecII}.

\subsection{Regimes} \label{Sec_Regimes}

We consider domains of size $32\ell_\mathrm{opt}\times 32\ell_\mathrm{opt}$ where $\ell_\mathrm{opt}\equiv2\pi/k_\mathrm{opt}$ is the optimal wavelength and depends on the $\Ra$ value used. In all the simulations the initial conditions are taken to be a small amplitude random field and the model equations are integrated for a sufficiently long time that a statistically stationary state is reached. All averages are calculated after discarding an initial transient.

\begin{figure}
	\centering
	\includegraphics[width=0.7\textwidth]{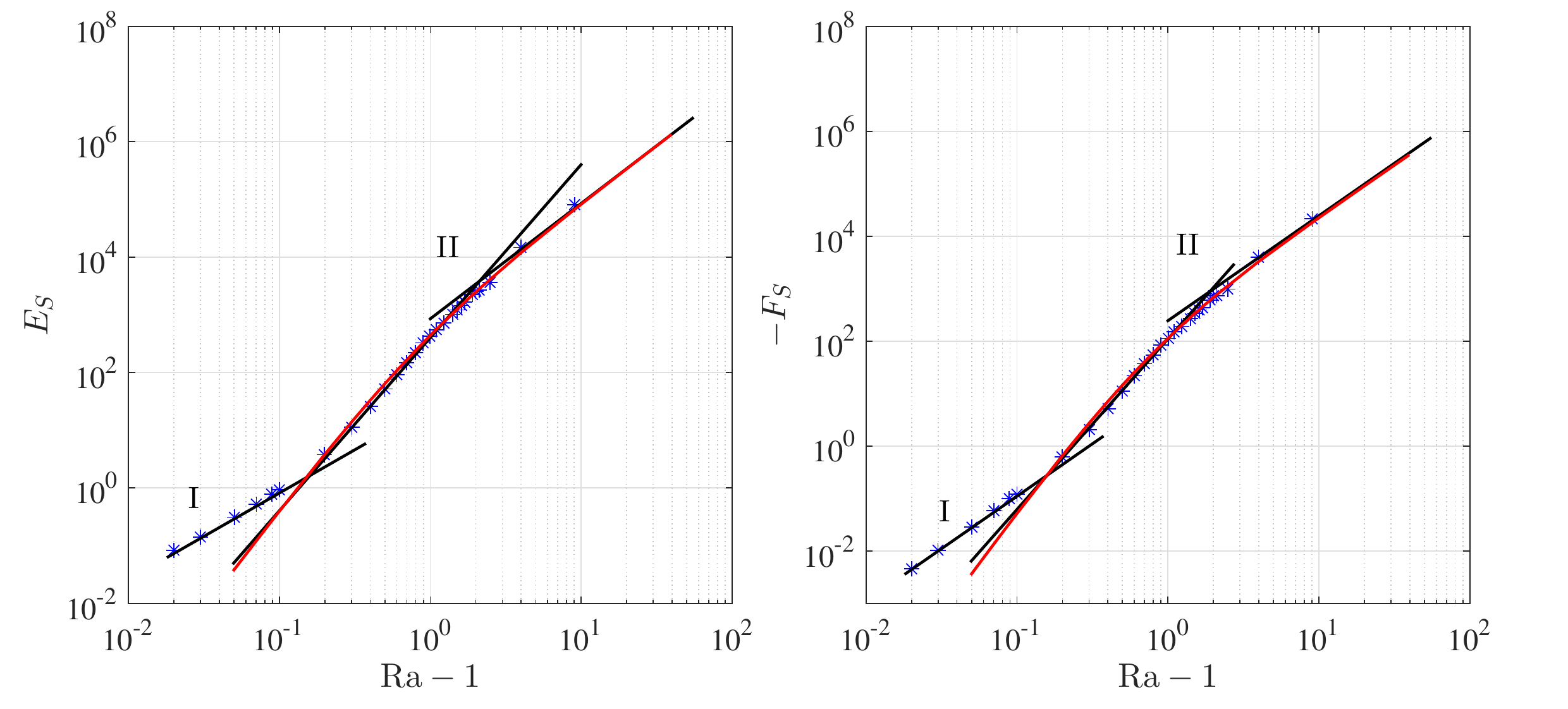}
	\caption{The dependence of $E_S$ and $-F_S$ on $\Ra$ in a log-log plot. Asterisks denote numerical results. The straight black lines have slopes $1.5$, $3$ and $2$ (left panel), and $2$, $3.25$ and $2$ (right panel). The red curves are the predictions (\ref{S_II}) and (\ref{psi_II}) valid in Regime II.}
		\label{Regimes}
	\end{figure}

In Fig. \ref{Regimes} we show the dependence of the energy $E_S\equiv(\half)\int S^2 \dd x \dd z/\int \dd x \dd z$ and the salinity flux $F_S\equiv -\int\psi_xS \dd x \dd z/\int \dd x \dd z$ on the parameter $\Ra$, ranging from $1.02$ to $10$. In this figure, three intervals of power-law dependence on the supercriticality $\R\equiv \Ra-1$ are seen. The first of these corresponds to Regime I, while the latter two correspond to different sublimits of Regime II.
	
\begin{figure}
		\centering
		\includegraphics[width=0.5\textwidth]{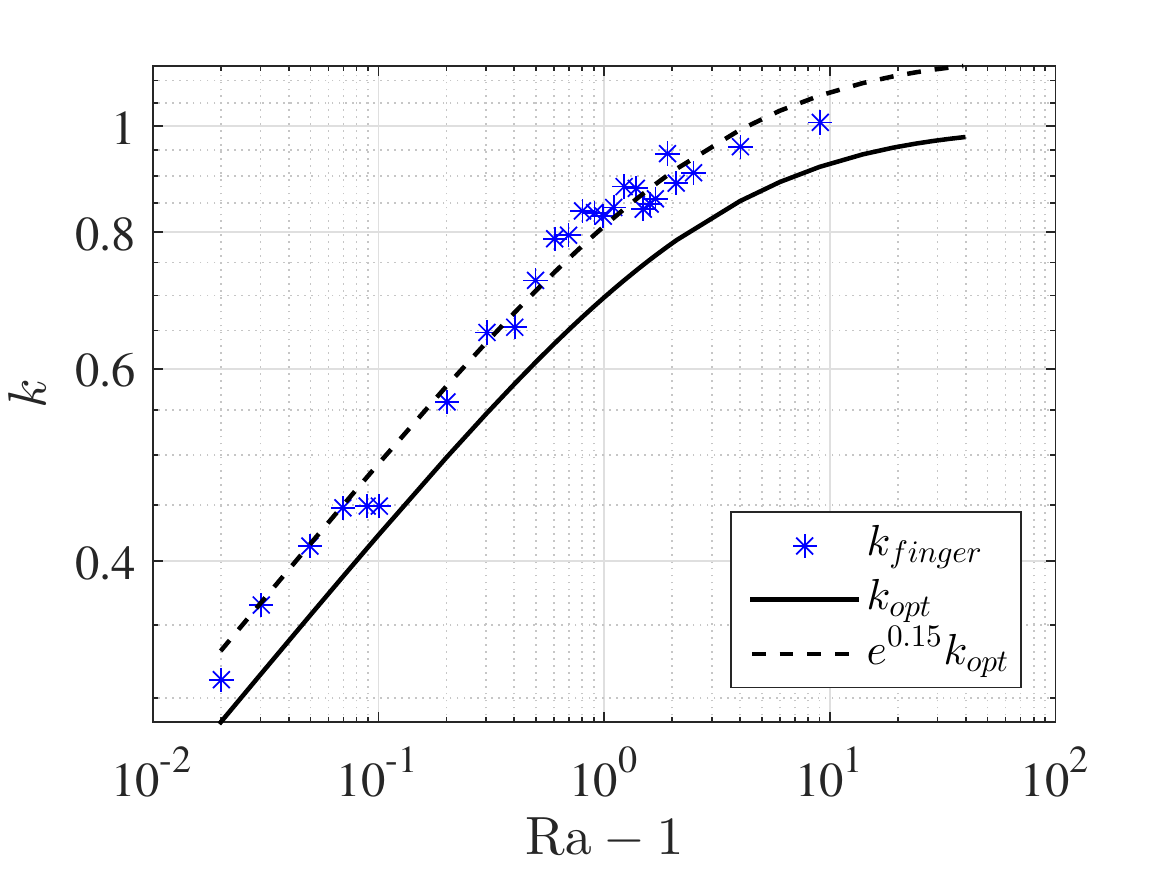}
		\caption{Horizontal finger scale vs $\Ra$. The quantity $k_\mathrm{finger}$ denotes the dominant wavenumber in the time-averaged spectrum of the statistically steady state while $k_\mathrm{opt}$ denotes the optimal wavenumber.}
		\label{Finger_scale}
	\end{figure}

	The dynamics of the IFSC model (\ref{RedMod}) are driven by the linear salt-finger instability, which has the distinctive feature that the optimal (horizontal) wavenumber is finite because of the coexistence of large and small scale damping stemming from the stabilizing temperature and viscosity, respectively. In the following we therefore assume that the characteristic horizontal scale of the system is determined by the optimal wavenumber $k_\mathrm{opt}$, and confirm this assumption in Fig. \ref{Finger_scale}, which shows the correspondence between the peak in the energy spectrum obtained from numerical simulations (Fig. \ref{Spectrum}) and expression (\ref{OptMode}) for the optimal wavenumber. Here the constant ratio $\ex^{0.15}\approx 1.16$ between $k_\mathrm{finger}$ and $k_\mathrm{opt}$ confirms the validity of our assumption. 
	
\begin{figure}
	\centering
	\includegraphics[width=0.8\textwidth]{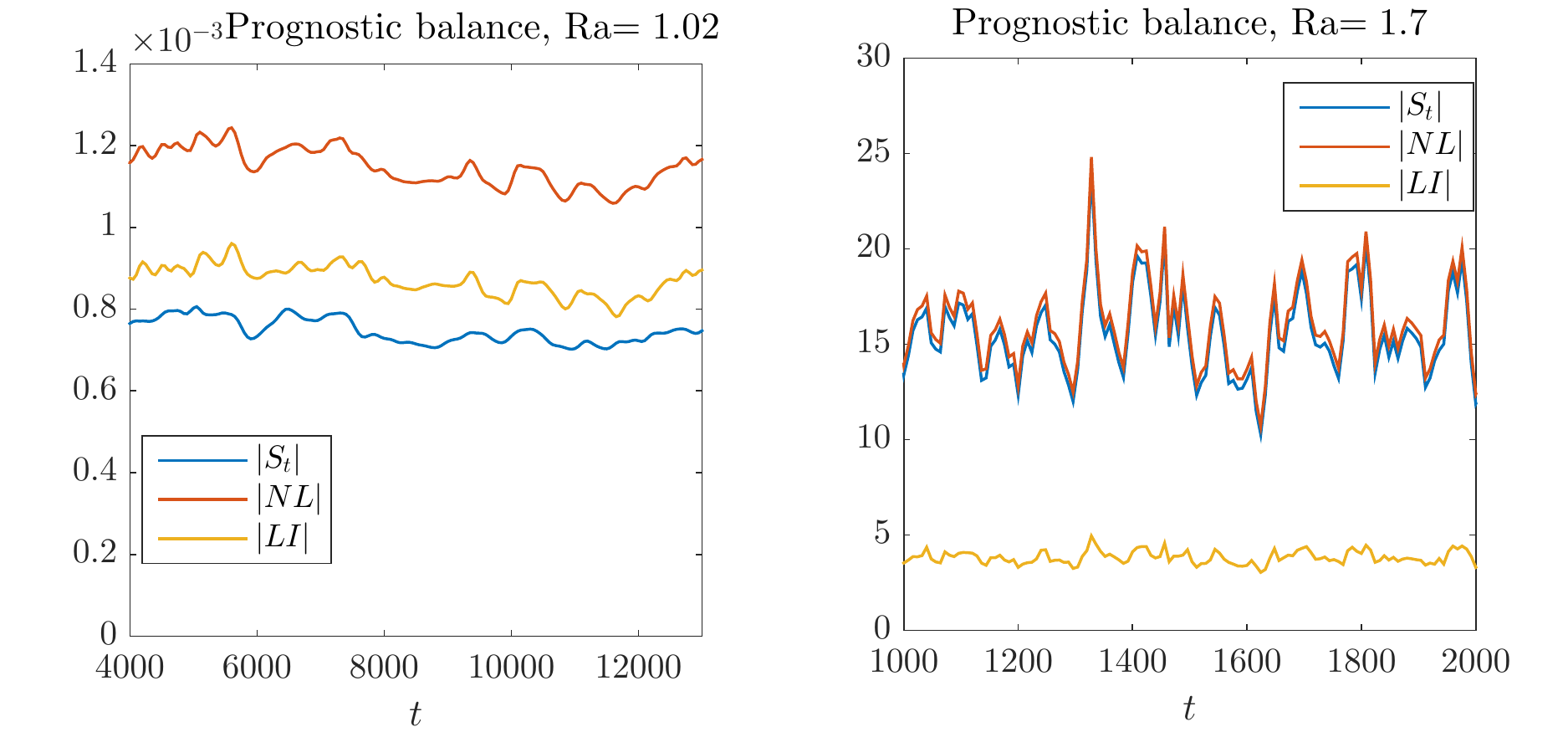}
	\caption{Balances in the prognostic equation in the statistically steady state when (a) $\Ra=1.02$ and (b) $1.7$. The quantities $S_t$, $NL$ and $LI$ are measured in terms of their r.m.s. values.}
	\label{Prognostic_balance}
\end{figure}

\begin{figure}
	\centering
	\includegraphics[width=0.9\textwidth]{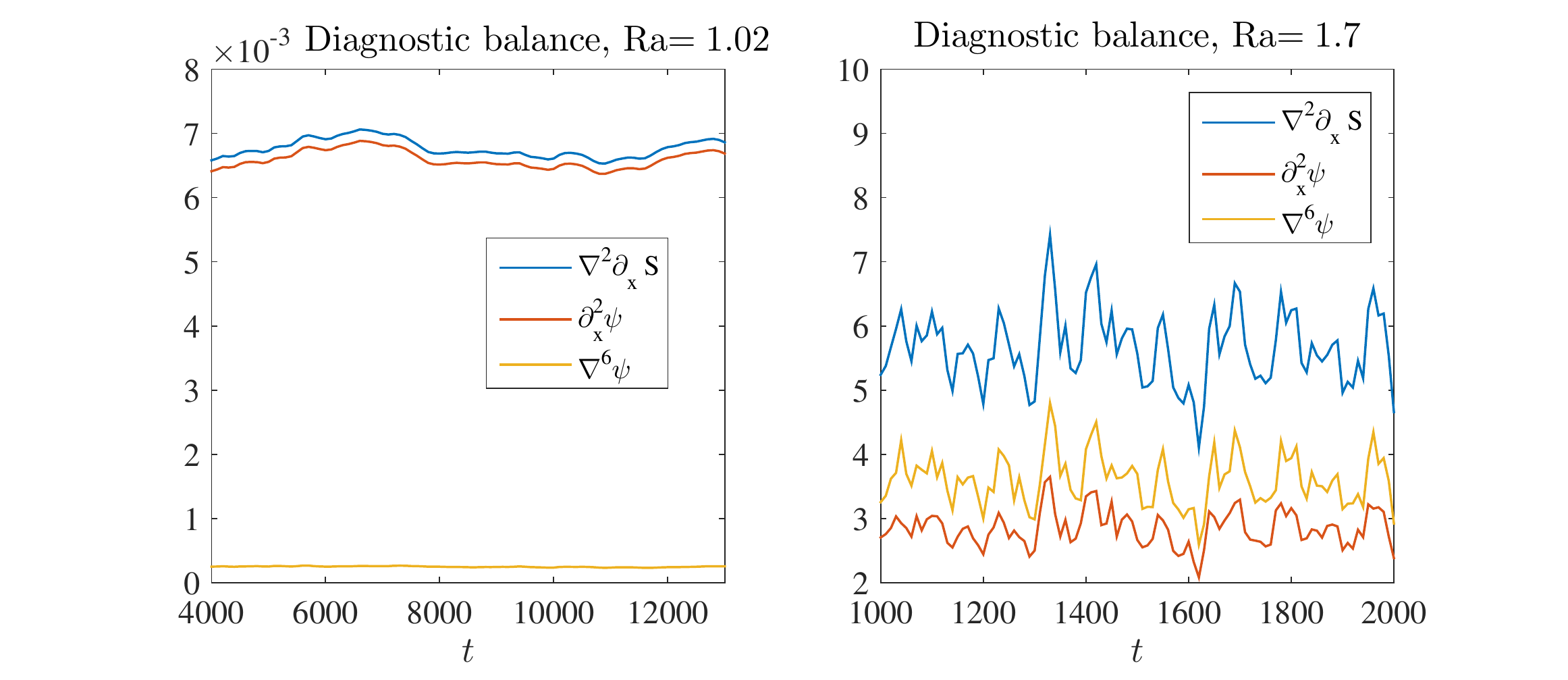}
	\caption{Balances in the diagnostic relation in the statistically steady state with $\Ra=1.02$, $1.7$, respectively. The quantities $S_t$, $NL$ and $LI$ are measured in terms of their r.m.s. values.}
	\label{Diagnostic_balance}
\end{figure}

	In Fig. \ref{Prognostic_balance} we compare the dominant balances in the prognostic equation with $\Ra=1.02$ and $1.7$. Here $NL=\Ja{\psi}{S}$ and $LI=\Ra\psi_x-\nabla^2S$ denote nonlinear advection and the linear part of the prognostic equation. In the left panel (Regime I) we find a balance between the time derivative, nonlinear advection and linear instability, while the right panel shows that Regime II is characterized by a balance between the time derivative and nonlinear advection only, with the small difference between these two terms balancing the remaining linear term $LI$. In Fig. \ref{Diagnostic_balance} we compare the corresponding dominant balances in the diagnostic relation at the same parameter values, $\Ra=1.02$ and $1.7$. In the left panel (Regime I) we find a dominant balance between $\pa{x}^2\psi$ and $\nabla^2\pa{x}S$, while in the right panel $\nabla^6 \psi$ becomes comparable to or larger than $\pa{x}^2\psi$ and so participates fully in the dominant balance.
The different dominant balances in the prognostic and diagnostic equations characterizing regimes I and II indicate that these regimes are fundamentally different. Specifically, consideration of the regime implications on the inertial-free diagnostic balance indicates that Regime I is dominated by a balance between the thermals and salinity buoyancy forces on the characteristic spatial scale of the fingers; recall that in Regime I $k_\mathrm{opt} \ll 1 $ (Fig. \ref{Opt_lambda_k}). In this regime viscous dissipation is only important at scales smaller than the finger scale and takes place primarily at the interface between fingers. Viscous dissipation gains importance in the inertia-free balance in Regime II. This occurs as a direct consequence of (i) the reduced separation between the fingering scale $2\pi/k_\mathrm{opt}\sim 1$  and the viscous dissipation scale, and (ii) the increased collisions between rising and descending fingers.
	
Based on the expression (\ref{OptMode}) for the optimal wavenumber and the dominant balances identified above, we can calculate the parameter dependences in both regimes.

\subsubsection{Regime I}

Regime I is a small supercriticality regime with $\R=\Ra-1\ll 1$, where the optimal scale is of $O(\R^{-1/4})$ (cf. Eq.~(\ref{OptMode})) and hence large. Here, by large scale we mean that the dominant balance in the diagnostic equation (\ref{Dia_Rel}) is 
\begin{equation}
\pa{x}^2 \psi \sim \nabla^2\pa{x}S, \label{Diag_I}
\end{equation}
a conclusion confirmed in the left panel of Fig. \ref{Diagnostic_balance}.  This statement implies an inviscid inertia-free balance between the thermal and salinity buoyancy forces. In this regime the appropriate balance in the prognostic equation (\ref{Pro_Rel}) is between the nonlinear and linear terms (Fig. \ref{Prognostic_balance}):
\begin{equation}
\Ja{\psi}{S} \sim -\Ra\psi_x+\nabla^2S \sim \lambda S, \label{Prog_I}
\end{equation}
where the last relation comes from the linear stability problem. Here $\lambda\sim \R^{3/2}$ as obtained from Eq.~(\ref{Opt_growth_rate}) and confirmed in Fig. \ref{Opt_lambda_k}.

We assume power-law dependence on $\R$ of the field magnitudes:
\begin{equation}
\psi\sim \R^a \quad \mathrm{and} \quad S \sim \R^b. \label{Assume_order_I}
\end{equation}
With the assumption that the order of magnitude of both the horizontal and vertical derivatives is determined by the optimal wavenumber, i.e.,
\begin{equation}
\pa{x} \sim \pa{z} \sim  \R^{1/4}
\end{equation} 
(see Eq.~(26a)) we substitute (\ref{Assume_order_I}) into (\ref{Diag_I}) and (\ref{Prog_I}) to obtain
\begin{subequations}
	\begin{align}
	1/2+a &= 3/4 + b,\\
	1/2+a+b &= 3/2 + b,
	\end{align}
\end{subequations} 
implying that
\begin{equation}
a=1 \quad \mathrm{and} \quad b =3/4.
\end{equation}
These exponents are used to generate the Regime I black line in both panels of Fig. \ref{Regimes}. We observe a good match between the theoretical and numerical results.

\subsubsection{Regime II} \label{SecII}

In Regime II, the supercriticality $\R$ becomes larger than that in Regime I and therefore in the diagnostic equation (\ref{Dia_Rel}) a full balance obtains (cf. Fig. \ref{Prognostic_balance}):
\begin{equation}
(\pa{x}^2 + \nabla^6)\psi \sim \nabla^2 \pa{x} S. \label{Dia_II}
\end{equation}
However, in contrast to Regime I, the leading order prognostic balance now involves the time derivative and the nonlinear terms only, leaving the linear term subdominant. In this regime we suppose that energy is transferred from the small energy injection scale, which peaks at the optimal wavelength of the fingers, to large scales, a scenario inspired by the observed increase in energy with increasing scale (Fig. \ref{Spectrum}) and the multiscale operator present in the diagnostic equation. This balance in energy transfer follows the same phenomenology as the energy cascade in the isotropic turbulence, where the dissipative effect is also subdominant in the prognostic momentum equation but determines the energy flux. However, our system differs from isotropic turbulence by the presence of two distinct dominant scales in the energy transfer that participate in the nonlocal interactions in spectral space.

The large-scale field must satisfy the diagnostic relation with negligible small-scale damping:
\begin{equation}
\pa{x}^2 \psi^{(l)} \sim \nabla^2\pa{x}S^{(l)}, \label{Dia_II_l}
\end{equation}
where the superscript $(l)$ indicates the large-scale field. As to the prognostic balance, we assume that the large-scale field reaches a distinguished regime such that the nonlinear and linear term balance:
\begin{equation}
\Ja{\psi^{(l)}}{S^{(l)}} \sim -\Ra\psi^{(l)}_x+\nabla^2S^{(l)} \approx \R\nabla^2 S^{(l)}. \label{Prog_II_l}
\end{equation}
Here the last relation takes advantage of the large scale of the fields and the definition of the supercriticality $\R$.

As in Regime I, we continue to assume that the scales of the fields are controlled by the linear theory optimal mode. However, since $k_\mathrm{opt}$ is no longer small we need to keep the full expression for the optimal wavenumber in Regime II:
\begin{equation}
\pa{x} \sim \pa{z} \sim  k_\mathrm{opt}.  \label{Finger_Scaling_II}
\end{equation}
Using (\ref{Finger_Scaling_II}) we can now calculate the $\Ra$-dependence of the large-scale streamfunction and salinity from (\ref{Dia_II_l}) and (\ref{Prog_II_l}):
\begin{equation}
\psi^{(l)} \sim \R \quad \mathrm{and} \quad S^{(l)} \sim \frac{\R}{k_\mathrm{opt}}. \label{Scaling_II_l}
\end{equation}  

We express the above-mentioned energy transfer picture in terms of a balance between the advection of potential energy ($S^2$) through small to large scale interaction and small scale energy pumping:
\begin{equation}
\Ja{\psi^{(\alpha)}}{S^{(\beta)}S^{(\gamma)}} \sim \Ra \psi_x S, \label{LSI_II}
\end{equation}
where $\Ja{\psi^{(\alpha)}}{S^{(\beta)}S^{(\gamma)}}$ is the larger of $\Ja{\psi^{(l)}}{SS}$, $\Ja{\psi}{S^{(l)}S}$, $\Ja{\psi^{(l)}}{S^{(l)}S}$ and $\Ja{\psi}{S^{(l)}S^{(l)}}$, and the quantities without the superscript $(l)$ denote small-scale fields. We need to calculate the four possibilities and check that the computed exponentials ensure the correct small to large scale interaction, i.e., that the chosen small to large scale advection term used to balance the energy input is the largest of the four. After going through these four cases, we conclude that the right term is $\Ja{\psi}{S^{(l)}S^{(l)}}$. This choice of interaction term is confirmed in Fig. \ref{Spectrum}: the salinity spectrum peaks at both large and small scales, while the flux spectrum only peaks at the small scale, implying that advection is dominated by the small-scale velocity.

The resulting balance 
\begin{equation}
\Ja{\psi}{S^{(l)}S^{(l)}} \sim \Ra \psi_x S 
\end{equation}
gives us the $\Ra$ dependence of $S$:
\begin{equation}
S \sim \frac{k_\mathrm{opt}}{\Ra}\br{S^{(l)}}^2 \sim \frac{\R^2}{k_\mathrm{opt}\Ra}. \label{S_II}
\end{equation}
Substituting (\ref{S_II}) into (\ref{Dia_II}) now yields
\begin{equation}
\psi \sim \frac{k_\mathrm{opt}}{1+Ck_\mathrm{opt}^4}S \sim \frac{\R^2}{(1+Ck_\mathrm{opt}^4)\Ra}, \label{psi_II}
\end{equation}
where the quantity $C$ measures the degree of isotropy. Empirically, we find that $C=2$ is a good match, for reasons that remain to be explored. 
Expressions (\ref{S_II}) and (\ref{psi_II}) are used in Fig. \ref{Regimes} to generate the red curves.

Finally, making use of (\ref{S_II}) and (\ref{psi_II}) we can obtain the scaling in the two subcases of Regime II: when $\R$ is comparable to unity, $k_\mathrm{opt}\sim \R^{1/4}$ and $\Ra=1+\R \sim \R^{1/4}$ (cf. Appendix \ref{Sec_expII}), and expressions (\ref{S_II}) and (\ref{psi_II}) can be approximated by a power-law dependence on $\R$:
\begin{equation}
\psi \sim S \sim \R^{3/2}. \label{Regime_II_1}
\end{equation}
When $\Ra$ is large, Eqs. (\ref{S_II}) and (\ref{psi_II}) become
\begin{equation}
\psi \sim S \sim \R^{1}. \label{Regime_II_2}
\end{equation}
We refer to these two regimes as II$_1$ and II$_2$, respectively. Expressions (\ref{Regime_II_1}) and (\ref{Regime_II_2}) are used to generate the four black lines in Fig. \ref{Regimes} when $\R$ is not small.

We summarize the scaling results for regimes I and II in Table \ref{Table_scalings}. 
\begin{table}
	\centering
	\begin{tabular}{ l|c|c|c|c|c|c|c|c|}
		& $S$ & $\psi$ & $F_S$  \\ \hline
		
		Regime I & $3/4$ & $1$ & $2$  \\
		
		Regime II$_1$& $3/2$ & $3/2$ & $13/4$  \\
		
		Regime II$_2$& $1$ & $1$ & $2$  \\
		
	\end{tabular}
	\caption{Exponents of $\R\equiv\Ra-1$ in the power-law scalings of $S$, $\psi$ and the salinity flux $F_S$ in regimes I and subregimes II$_1$ and II$_2$ for small and large $\R$, respectively.}
	\label{Table_scalings}
\end{table}

	\subsection{Probability density functions} \label{Sec_pdf}
	
\begin{figure}
		\centering
		\includegraphics[width=\textwidth]{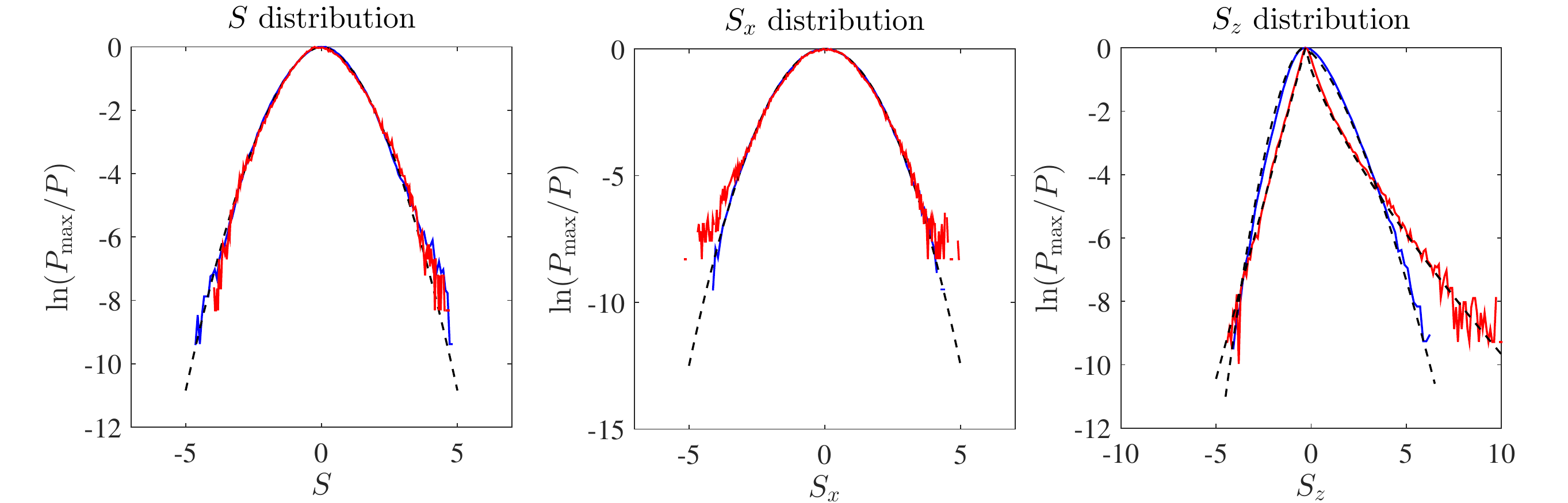}
		\caption{Pdfs of $S$, $S_x$ and $S_z$ for $Ra=1.02$ (blue) and $1.7$ (red). Coefficients of the dashed stretched-Gaussian curves shown alongside are given in Table \ref{pdf_Parameter}.}
		\label{pdf_S}
	\end{figure}

\begin{figure}
		\centering
		\includegraphics[width=\textwidth]{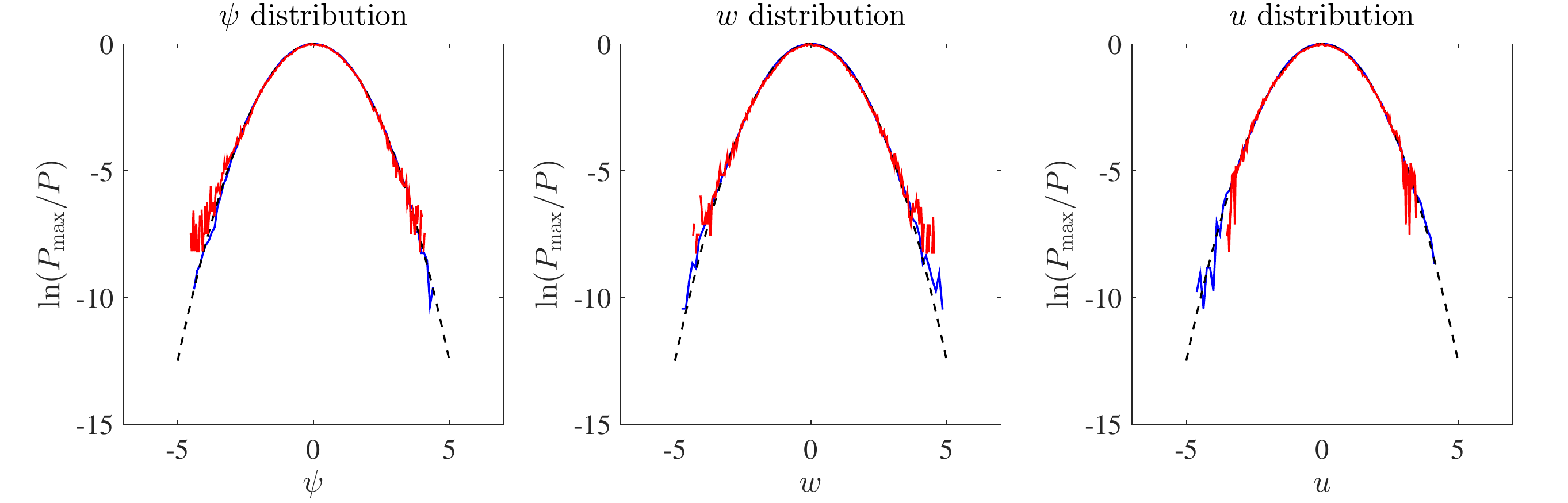}
		\caption{Pdfs of $\psi$, $w$ and $u$ for $Ra=1.02$ (blue) and $1.7$ (red). Coefficients of the dashed stretched-Gaussian curves shown alongside are given in Table \ref{pdf_Parameter}.}
		\label{pdf_psi}
	\end{figure}

	To obtain a detailed understanding of the statistics of the saturated fields, we now turn to the study of the probability density functions (pdfs) of the different fields. In Fig. \ref{pdf_S} and \ref{pdf_psi} we show the pdfs of the salinity and its derivatives and the pdfs of streamfunction and its derivatives of both Regime I ($\Ra=1.02$) and Regime II ($\Ra=1.7$). For $\Ra=1.02$, the pdfs are obtained from data over $t\in(12000,\,13000)$ with step size $\Delta t=100$, while for $\Ra=1.7$ we show one snapshot at $t=791$. 
	All pdfs presented are normalized by their variance.


	We observe that most fields obey Gaussian or quasi-Gaussian statistics. For more detailed assessment, we fit the pdfs to a general stretched-Gaussian form:
	\begin{equation}
	P(a)=P_\mathrm{max}\exp^{-\abs{a/\beta}^\alpha}, \label{pdf_form}
	\end{equation}
	where $\alpha$, $\beta$ and $P_\mathrm{max}$ are constants and $\alpha=2$ corresponds to the Gaussian distribution. The parameter $\beta$ is adjusted to obtain a probability density function with variance 1, and $P_{\mathrm{max}}$ normalizes the distribution:
	\begin{equation}
	P_\mathrm{max}=\frac{\alpha}{2\beta\Gamma(1/\alpha)} \quad \mathrm{and} \quad \beta^2 = \frac{\Gamma(1/\alpha)}{\Gamma(3/\alpha)}, \label{Coe_relations}
	\end{equation}where $\Gamma$ is the gamma function. From the second relation above we calculate $\beta$ in Table \ref{pdf_Parameter} for the pdfs of $S$, $S_x$, $\psi$, $w$ and $u$. This form has been utilized in the context of Rayleigh-B\'enard convection in \cite{Chin1991}.
	
	In addition to the above general form, there are several other interesting features that can be learned from these figures. In the right panel of Fig. \ref{pdf_S} the pdfs of $S_z$ are not symmetric with respect to their peaks, indicating obvious skewness. This asymmetry originates from the collisions between positive (rising) and negative (descending) fingers: collisions occur where $S_z$ is positive and tend to increase the salinity gradient. Nontrivial skewness is a common feature of convection where coherent structures such as fingers or plumes exist, and has been studied in Rayleigh-B\'enard convection to identify plume generation \cite{Belmonte1996}. However, there is no such asymmetry in $\psi$ because $\psi_z$ is the horizontal velocity, which preserves on average the reflection symmetry $u\to -u$ of the system. The pdfs of $S_x$, $\psi$, $w$ and $u$ are all Gaussian (c.f. Fig. \ref{pdf_S} and \ref{pdf_psi}), which is very useful for parameterization. Since \S \ref{Sec_Regimes} develops a theory for the dependence of the variance of the different fields on the parameter $\R$ (equivalently $\Ra$), the statistics of the quantities with Gaussian distribution are fully understood, and these can be used parameterize the diffusivity induced by salt-fingering convection as $\Ra$ varies.
	
	\begin{table}
		\centering
		\begin{tabular}{|c|c|c|c|c|}
			\multicolumn{1}{c}{} & \multicolumn{2}{|c|}{$\Ra=1.02$} & \multicolumn{2}{|c|}{$\Ra=1.7$} \\
			\cline{2-5}
			\multicolumn{1}{c|}{} & $\alpha$ & $\beta$ &  $\alpha$ & $\beta$ \\ \hline
			$S$ & 1.8 & 1.33  & 1.8 & 1.33   \\
			$S_x$ & 2 & 1.41 & 2 & 1.41  \\
			$S_z^+$ & 1.5 & 1.43  & 0.75 & 0.5  \\
			$S_z^-$ & 1.7 & 1  & 1 & 0.45  \\
			$\psi$ & 2 & 1.41  & 2 & 1.41  \\
			$w$ & 2 & 1.41  & 2 & 1.41  \\
			$u$ & 2 & 1.41  & 2 & 1.41  \\
		\end{tabular}
		\caption{Curve-fitting parameters for the probability density functions in Figs. \ref{pdf_S} and \ref{pdf_psi}.Here the $``+"$ and $``-"$ symbols denote the positive and negative arguments of pdfs, respectively.}
		\label{pdf_Parameter}
	\end{table}

\section{Discussion} \label{Sec_Dis}

In this paper, we systematically derived two reduced models for salt-fingering convection in the limit of small diffusivity and large density ratios, $\tau \ll 1$ and $R_\rho=O(\tau^{-1})$. In the modified RBC model (\ref{mRBC}) the stabilizing temperature plays the role of large-scale damping, which when combined with the small-scale viscosity leads to a finite optimal wavenumber corresponding to the fastest linear growth rate. This model may find application in the astrophysical context where $\Sc=O(1)$. In the parameter regime corresponding to the oceans where $\Sc$ is large we obtained a simpler model, the inertia-free salt convection model (\ref{RedMod}), which has a prognostic-diagnostic form, and we studied this model in detail. In large domains, corresponding to large thermal and solutal Rayleigh numbers, we checked that this model preserves the linear and secondary instabilities present in the primitive equations and that it captures the three stages -- finger dominance, finger disruption and saturation -- that have been observed to lead to a statistically steady saturated state. Both secondary instability and finger collision were found to be responsible for the presence of this state, thereby extending the conclusion reached by \citet{Shen1995} that finger collisions are important at $O(1)$ density ratios to the limit of large density ratio.

The properties of the saturated state depend on the Rayleigh ratio $\Ra$. We identified two parameter regimes characterized by different balances in both the diagnostic and prognostic equations, corresponding to weak and strong supercriticalities, leading to distinct saturated states. In Regime I, where the width of salt finger scale as $k_\mathrm{opt}^{-1}\sim \R^{-1/4}\ll 1$, the temperature and salinity buoyancy effects dominate the diagnostic equation and viscous dissipation is negligible. In the prognostic equation, a balance between linear instability and nonlinear advection indicates that the growth rates of primary and secondary instabilities are of the same order (cf. \cite{Radk2003,Brown2013}). In this regime, we find power-law dependences on the supercriticality of the saturated state. Regime II occurs at larger supercriticality. Even though this indicates stronger driving forces, viscous dissipation enters the diagnostic balance. This counterintuitive fact results from the smaller salt-finger scale at larger supercriticalities. In addition to the small salt-finger scale, a large scale is also present. This scale is fed by an energy transfer from the small finger scale at which energy is injected into the system and it is this energy transfer that characterizes Regime II. The result relies on the dependence of the energy injection scale on the Rayleigh ratio, leading to two sublimits with distinct power-law dependence.

The presence of a saturated state even in the presence of exact elevator mode solutions that grow exponentially forever requires that our simulations are initialized with broad-band initial conditions, cf. \cite{Calzavarini2006,Jamroz2008}. In this case the system is unable to `find' the growing elevator mode state resulting in a saturated state exhibiting a broad range of spatial scales. Figure \ref{Energy_field}(a) shows an example of the transient generation of elevator modes followed by their disruption via secondary instability. The resulting statistically steady state is characterized by pdfs that have in general a stretched Gaussian form. In particular, the horizontal and vertical velocities have Gaussian distributions and can therefore be used to calculate the diffusivity of turbulent salt-finger convection, which can be further developed into a parameterization of transport processes. In this connection we mention that in two-dimensional Rayleigh-B\'enard convection such Gaussian pdfs are universal \cite{Celani2002}. Moreover, passive tracers in such a system have identical statistics \cite{Celani2002}. In salt-finger convection it is the salinity field that drives convection, but a tracer like the temperature is not passive since it contributes to buoyancy. As a result one might expect departures from Gaussian pdfs and indeed such departures have been detected \cite{Hardenberg2010}. However, in both our reduced models the temperature field is slaved to the streamfunction and the resulting models resemble the equations governing two-dimensional Rayleigh-B\'enard convection. We conjecture that this is the reason why we find Gaussian velocity pdfs in our computations. 

Our IFSC model (\ref{RedMod}) is linked to the weakly nonlinear model derived by \citet{Radk2010} (see Eqs. (8) and (11) of \cite{Radk2010}). Even though different parameter regimes are considered in the two models --  our reduced model for $\tau\to 0$ captures a fully nonlinear regime where dynamics are not confined to the onset of instability ($\Ra$ ranges from $1$ to $\infty$), while Radko's \cite{Radk2010} model relaxes the small $\tau$ assumption but is restricted to dynamics near the onset -- the two models match in the overlapping regime $\tau\ll 1$ and $\Ra=1+\epsilon {R}$ with $\epsilon\ll 1$ and ${R}=O(1)$. In this regime, the optimal scale from Eq.~(\ref{OptMode}) is of $O(\epsilon^{-1/4})$, suggesting the rescaling
\begin{equation}
\pa{t} \to \epsilon^{9/4}\pa{t}, \quad \nabla\to\epsilon^{1/4}\nabla, \quad \psi \to \epsilon\psi \quad \mathrm{and} \quad S \to \epsilon^{3/4} S. 
\end{equation}
With this rescaling the leading order contribution to (\ref{RedMod}) reads
\begin{equation}
\pa{t}S + \Ja{\pa{x}^{-1}\nabla^2S}{S} + \br{{R}-\pa{x}^{-2}\nabla^6}\nabla^2 S = 0, \label{RRedMod}
\end{equation}
which is identical to the model derived by \citet{Radk2010} in the limit of small $\tau$.

In our simulations we found that the characteristic length of the fingers increases as $\Ra$ increases (cf. Figs. \ref{Energy_field} and \ref{Field}), in contrast to a result of Merryfield and Grinder (cf. Fig. 8 in \cite{Youshida2003}). Evidently salt-finger convection exhibits different limiting parameter regimes: in our model $\tau$ and $\Rr^{-1}$ are constrained to be of the same order and to be small, while in the simulations of Merryfield and Grinder the value of $\tau$ is fixed. 


We may define a salinity Nusselt number by 
\begin{equation}
\Nu_S =- \frac{F_S}{\Ra},
\end{equation}
and use the Regime II scalings to obtain $\Nu_S\!\!\sim\!\!\Ra$ in the limit of large $\Ra$. This result is to be compared with the asymptotic result $\Nu_S\!\!\sim\!\! \Ra^{1/3}$ obtained by \citet{Yang2015} for a vertically bounded domain. Evidently the presence of elevator modes permitted in the doubly periodic domain leads to a very substantial enhancement of the salinity flux. In view of (\ref{RaT}) our thermal Rayleigh number $\Ra_T$ is of order $10^7$ while the salinity Rayleigh number $\Ra_S$ is of order $10^8$ (see (\ref{DefRa})). These values are comparable to those used in \cite{Yang2015} although our density ratio is much larger than that in \cite{Yang2015}.

Salt-finger convection often results in the formation of salinity staircases \cite{Schmitt2005,Radk2003,Stellmach2011,Paparella2012}, and these have been extensively studied in oceanographic measurements, numerical simulations and theories. In our IFSC model we have not observed the formation of staircases. This may be because our model filters out gravity waves, which are believed to be important for staircase formation through collective instability \cite{Ster1969,Holy1984}. The $\gamma$-instability mechanism proposed by \citet{Radk2003} provides an alternative explanation but requires a nonmonotonic dependence of the salt flux on the mean salinity gradient. Since in our system this dependence is always monotonic neither mechanism for staircase formation is present. However, \citet{Brown2013} recently discovered that staircases can form even when the flux increases monotonically with the mean salinity gradient raising the possibility that the Rayleigh ratio used in our simulations is insufficiently large to generate this interesting state. 

In a future publication we plan to discuss the extent to which the conclusions of this paper carry over to three dimensions.

\vspace{6pt} 


{\bf This work was supported in part by the National Science Foundation under grant DMS-1317596 (JHX and EK) and under grant DMS-1317666 (KJ and BM). We thank Pascale Garaud for helpful discussions.}

\appendix

\section{Domain-size dependence} \label{Sec_domain_size}

In this Appendix we briefly study the effect of the domain size on the saturation of the salt-finger instability when $\Ra=1.1$. In the reduced model, the length normalization indicates that larger domain heights represent larger Rayleigh numbers for both $S$ and $T$ fields (cf. Eq. (\ref{RaT})), and that fixed $\Ra=\Ra_S/\Ra_T$ indicates a constant ratio between them. For this $\Ra$ the optimal scale is of $O((\Ra-1)^{-1/4})$ (cf. Eq. (\ref{OptMode})) which is large. 
Here we study small and intermediate domains with domain size of $1\times 2$ and $8\times 8$ times the optimal wavelength $\ell_\mathrm{opt}=2\pi/k_\mathrm{opt}$. In all the simulations initial conditions are taken to be small amplitude Gaussian random fields.

\begin{figure}
	\centering
	\includegraphics[width=0.9\textwidth]{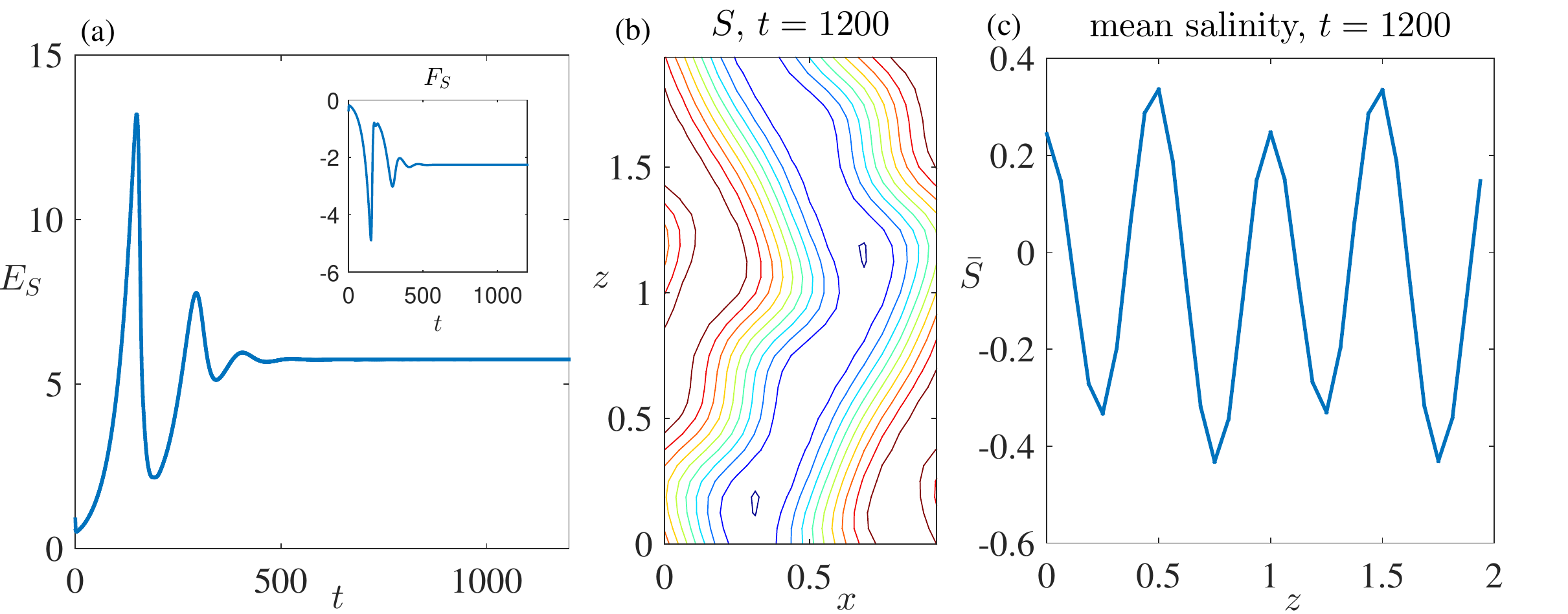}
	\caption{Results from a $\ell_\mathrm{opt}\times 2\ell_\mathrm{opt}$ simulation for $\Ra=1.1$. (a) Evolution of the total energy with time. (b) Large-time salinity field. (c) Associated horizontal mean salinity. Distances are measured in units of the wavelength of the optimal mode.}
	\label{1time2}
\end{figure}

In Fig. \ref{1time2}, we show the dynamics in a $1\times 2$ domain. Panel (a) shows that the time evolution of total energy $E_S$. The total energy increases almost exponentially after a very brief decrease caused by the damping of stable modes in the initial state. This exponential growth corresponds to the linear instability of the optimal mode. The secondary instability suppresses the growth of the instability leading to decaying oscillations in the energy $E_S$ before the system finally reaches a steady state. Properties of this final state are shown in panels (b) and (c) in terms of the salinity field at large times and the corresponding horizontal mean flow, respectively, and these resemble similar results obtained in Ref. \cite{SternRadko1998}.

\begin{figure}
	\centering
	\includegraphics[width=\textwidth]{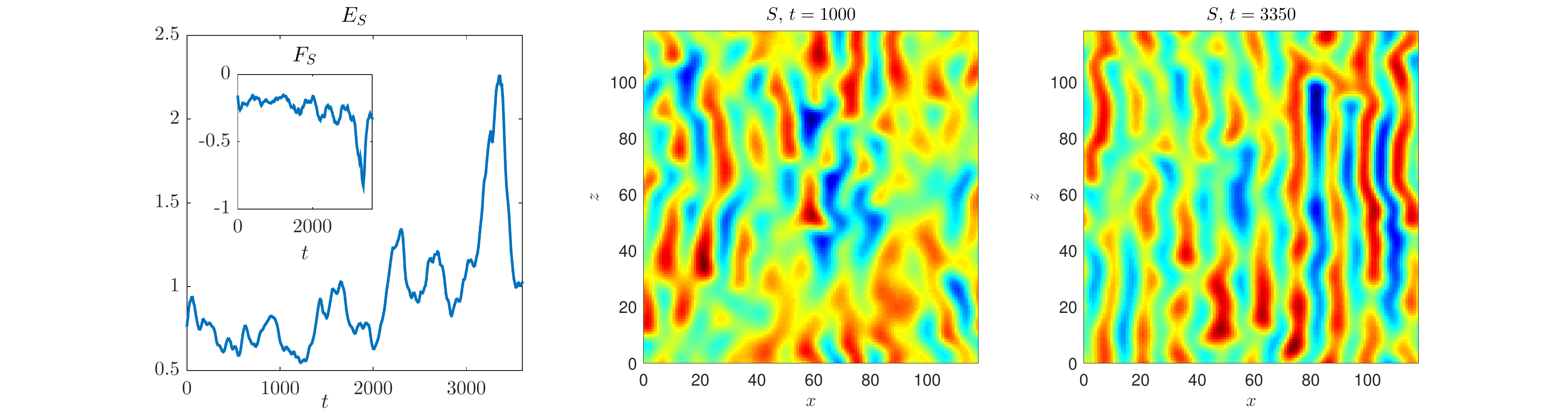}
	\caption{Results from an $8\ell_\mathrm{opt}\times 8\ell_\mathrm{opt}$ simulation for $\Ra=1.1$. The left panel shows the time evolution of total energy and vertical salinity flux. The middle and right panels show the instantaneous salinity fields at $t=1000$ and $3350$, respectively.}
	\label{ES_FS_88}
\end{figure}

With increasing domain size the system becomes chaotic, much as observed in \citet{Trax2011}. In Fig. \ref{ES_FS_88} we show such chaotic behavior in a simulation in a domain of size $8\times 8$ the optimal mode wavelength. The left panel reveals the presence of bursts in both energy and flux. As shown in the middle and right panels the peaks arise when transport is dominated by salt fingers. When the salt fingers are disrupted by the secondary instability the peaks subside. Since the growth rate of secondary instabilities depends on the amplitude of the salt fingers, disruption of the salt finger field suppresses the secondary instability, allowing the salt finger field to regrow. There is thus a natural mechanism that is responsible for presence of oscillations between organized and disordered convection. Related mechanisms are responsible for the presence of relaxation oscillations in binary fluid convection \cite{Batiste2006} and in Rayleigh-B\'enard convection in an imposed magnetic field \cite{Matthews1993}. 
These results are to be compared with the $32\ell_\mathrm{opt}\times 32\ell_\mathrm{opt}$ results in the body of the text.

\section{Exponential approximation in Regime II} \label{Sec_expII}

Fig. \ref{Regimes} indicates that Regime II contains two sublimiting regimes where the magnitudes of the fields have power-law dependence on the supercriticality $\R$. 
To obtain the expressions for the sublimiting Regime II$_1$ from (\ref{S_II}) and (\ref{psi_II}), we introduce a useful approximation $\Ra=1+\R \sim \R^{1/4}$. 
Here, the appropriate choice of the exponent, $\alpha=1/4$, is determined by minimizing the error of this approximation, which is defined as 
\begin{equation}
e_\mathrm{Ra} = \max\left\{ \ln \br{ \frac{\Ra}{\R^{\alpha}} } \right\} -  \min\left\{ \ln \br{ \frac{\Ra}{\R^{\alpha}} } \right\}.
\end{equation}
Here the maximum and minimum are taken over the interval $\Ra-1\in\sbr{\ex^{-2},\,1}$ corresponding to Regime II$_1$ (cf. Fig. \ref{Regimes}). 

\begin{figure}
	\centering
	\includegraphics[width=0.5\textwidth]{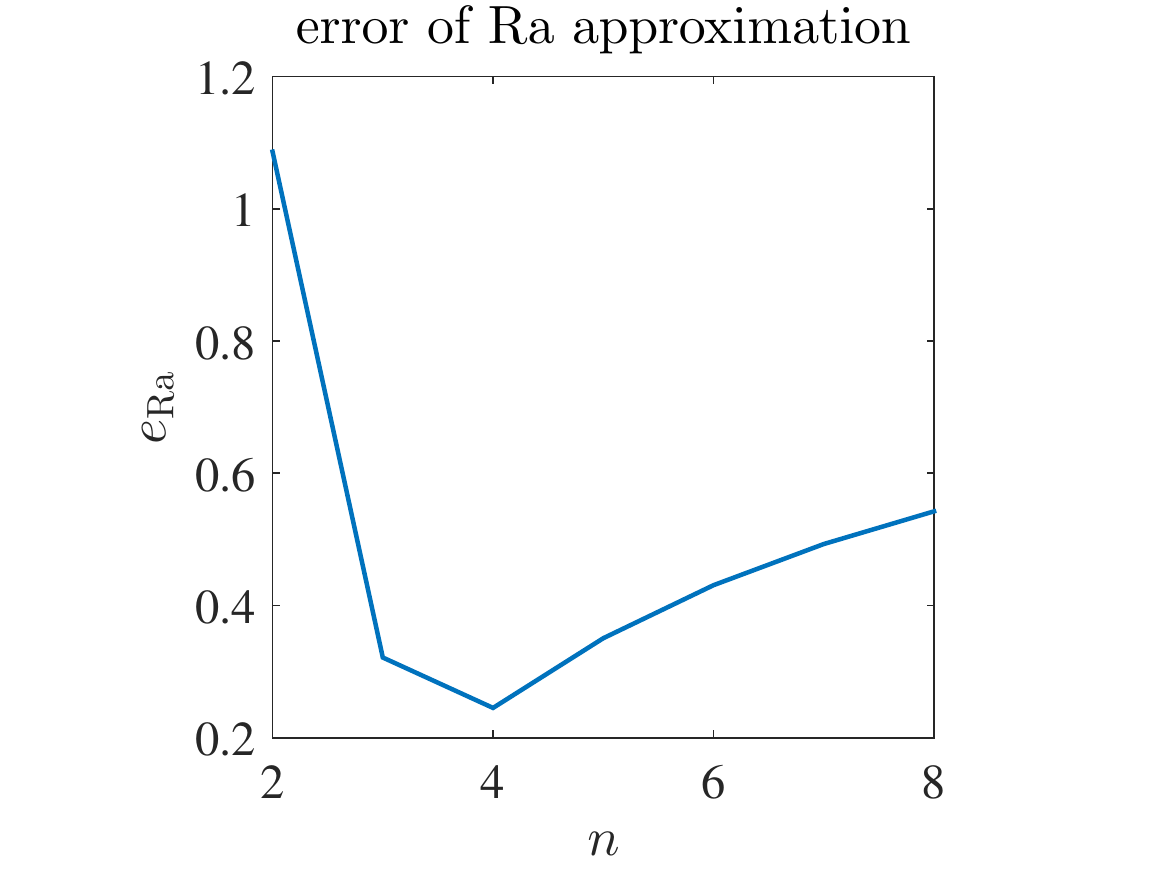}
	\caption{Error in the power-law approximation of $\Ra$ as a function of exponentials. Here $n=1/\alpha$ is chosen to be an integer: $n=2,\,3,\,4,\,5,\,6,\,7,\,8$. }
	\label{error_Ra}
\end{figure}

Fig. \ref{error_Ra} shows the error $e_\mathrm{Ra}$ as a function of $n=1/\alpha$ with a minimum at $n=4$, justifying our choice of the exponent $\alpha=1/4$.  
This approximation brings about an error of $1/10$ to the exponent $\alpha$ -- $\Ra\sim \R^{1/4\pm 1/10}$,
and therefore brings about a difference of $1/10$ when calculating the exponents of $\psi$ and $S$ from (\ref{S_II}) and (\ref{psi_II}), but this difference is negligible compared with $3/2$, the exponents of $\psi$ and $S$. 
Instead of $\alpha=1/4$, if we choose $\alpha=1/3$ or $\alpha=1/5$ the calculated exponents of $\psi$ and $S$ will deviate from $3/2$ by differences of $1/12$ or $1/20$, which are also negligible. 
Therefore we take the optimal choice of the exponent $\alpha=1/4$ and use it to generate the black lines for Regime II$_1$ in both panels of Fig. \ref{Regimes}. 
We need to mention that the power-law expressions for Regime II$_1$ are only numerical approximations which are not asymptotic.



\bibliographystyle{mdpi}




\end{document}